\documentclass[prb,a4paper,showpacs,twocolumn,
]{revtex4}
\usepackage{amsmath}
\usepackage{amssymb}
\usepackage{amsfonts}
\usepackage{graphicx}
\usepackage{epstopdf}
\usepackage{color}
\usepackage{bm}

\DeclareMathOperator{\Tr}{Tr} 
\DeclareMathOperator{\tr}{tr}
\DeclareMathOperator{\sgn}{sgn}

\DeclareMathOperator{\erfi}{erfi}
\DeclareMathOperator{\erf}{erf}
\newcommand{\ve}{\varepsilon}
\newcommand{\nup}{n_{\uparrow}}
\newcommand{\ndn}{n_{\downarrow}}

 1

\begin{document}

\title{An Exact Solution for Spin and Charge Correlations in Quantum Dots: \\
The Effect of Level Fluctuations and Zeeman Splitting}

\author{I.S. Burmistrov$^{1}$, Yuval Gefen$^{2}$ and M.N. Kiselev$^{3}$}
\affiliation{$^{1}$ L.D. Landau Institute for Theoretical Physics RAS,
Kosygina street 2, 119334 Moscow, Russia}
\affiliation{$^{2}$ Department of Condensed Matter Physics, The Weizmann Institute of Science, Rehovot 76100, Israel}
\affiliation{$^{3}$ International Center for Theoretical Physics, Strada Costiera 11, 34014 Trieste, Italy}

\begin{abstract}
The inclusion of charging and spin-exchange interactions within the
Universal Hamiltonian description of quantum dots  is challenging as
it leads to a non-Abelian action.  Here we present  an {\it exact}
analytical solution of the probem, in particular, in the vicinity of
the Stoner instabilty. We calculate the tunneling density of states and the spin
susceptibility. We demonstrate that near the Stoner instability the spin susceptibility follows a Curie law with an effective spin. The latter
depends logarithmically on temperature due to the statistical fluctuations of the single-particle levels. Near the Stoner instability the tunneling density of states exhibits a non-monotonous behavior as function of the tunneling energy, even at
temperatures higher than the exchange energy. This is due to ehnanced spin correlations. Our results could be tested in quantum dots made of nearly ferromagnetic
materials.
\end{abstract}


\date{\today}
\maketitle

\section{Introduction\label{Sec.Intro}}

The study of quantum dots (QDs) is at the cutting edge of modern condensed matter physics. 
The introduction~\cite{ABG,KAA}  of the universal Hamiltonian (UH) made it possible under not severe assumptions to describe a variety of QDs by means of an effective zero-dimensional Hamiltonian with a few physical parameters. In particular, the UH allows one to simplify the intricate electron-electron interactions within a QD in a controlled way. 

Within the framework of the UH electron-electron interaction is represented as the sum of three spatially independent terms: charging (singlet particle-hole channel), spin-exchange (triplet particle-hole channel), and  interaction in the Cooper channel. The latter is responsible for superconducting correlations in QDs. In what follows, we shall assume that the Cooper channel is suppressed, e.g., by the orbital effect of a weak magnetic field. The charging term is responsible for the well-known phenomenon of Coulomb blockade.~\cite{CB} It is broadly known that in the presence of significant ferromagnetic spin-exchange interaction, bulk systems can undergo a Stoner transition from para- to ferromagnetic materials. In the case of QDs physics is richer.~\cite{KAA} One distinguishes three regimes of behavior as function of the increased strength of the ferromagnetic exchange interaction ($J>0$): i) paramagnetic (the total spin in the ground state is zero); ii) mesoscopic Stoner regime (finite total spin in the ground state whose value increases stepwise with the exchange); and iii) thermodynamic ferromagnetic phase (the total spin in the ground state is proportional to the volume of a QD). The mesoscopic Stoner regime disappears in the thermodynamic limit: $\delta\to 0$, where $\delta$ is the mean spacing between single-particle energy levels in a QD. The mesoscopic Stoner regime is sensitive to the statistical fluctuations of single particle levels~\cite{KAA} and to the presence of the Zeeman splitting.~\cite{Schechter}  The former enhances the total spin in the ground state whereas the latter suppresses the mesoscopic Stoner instability. To take both effects into account simultaneously calls for a full-fledged quantum mechanical treatment of the problem.

At first glance, the UH with charging and spin-exchange interaction terms is easy to solve. All its three parts (free electron term, charging term, and spin-exchange term) commute with each other. It allows one to work in a basis of states classified by  the total number of electrons and the total spin.~\cite{AlhassidRupp,AlhassidRuppKaminskiGlazman,AlhassidTureci} However, this approach requires calculation of Clebsch-Gordan coefficients which is not an easy task. In this way Alhassid and Rupp~\cite{AlhassidRupp} have found an exact solution for the partition function in the absence of Zeeman splitting. Elements of their analysis were then incorporated into a master equation analysis of electric~\cite{AlhassidRupp,AlhassidRuppKaminskiGlazman} and thermal~\cite{AlhassidStone} transport through a QD at low temperatures $T\lesssim\delta$. Independently, a study of electron transport through a QD for the same temperature range, $T\lesssim\delta$, was made by Usaj and Baranger.~\cite{UsajBaranger} Their analysis, accounting for the charging and exchange interactions, was based on a master equation approach as well.

More traditional way to tackle an interacting problem is to employ the Hubbard-Stratonovich transformation. The latter reduces the interacting problem to the problem of non-interacting electrons in the presence of dynamical (time-dependent) external field (see e.g., Ref.~[\onlinecite{ZaikinSchoen}]. For the case of the UH with the charging interaction only such dynamical external field can be removed by suitable gauge transformation of fermionic operators. This method was first employed by Kamenev and Gefen~\cite{KamenevGefen} and resulted in the exact solution of the problem.~\cite{EfetovTscherisch,SeldmayrLY} The gauge transformation can also be treated by means of stochastic (Langevin) equations. Corresponding stochastic
bosonization approach based on construction and solution of Fokker-Plank
equations was used in Ref.~[\onlinecite{Boaz}] to solve exactly the UH with the charging and
Ising-spin-exchange interactions.

In the case of isotropic spin-exchange interaction the Hubbard-Stratonovich transformation results in the effective action which describes electrons with a dynamical time-dependent magnetic field $\bm{\theta}(t)$ acting on their spins.~\cite{KG}  Solving this non-Abelian effective action is an intricate problem. One needs to tackle time ordered exponents of the form
\begin{equation}
\label{eq_Intro}
\mathcal{T} \exp \left ( i \int_0^{t} dt^\prime\, \bm{\theta}(t^\prime) \bm{s}(t^\prime) \right )  ,
\end{equation}
where $\bm{s}$ represents an electron spin and $\mathcal{T}$ is a time ordering operation. To avoid the problem of the time ordering the UH with anisotropic spin-exchange interaction was considered, and perturbation expansion around Ising point was performed.~\cite{KG}  

Here we present an {\it exact} analytic algorithm to tackle the challenging problem of the UH with charging and isotropic spin-exchange interactions in the presence of Zeeman spliting. To solve the problem of a time ordered exponent we employ here  a
Wei-Norman-Kolokolov (WNK) transformation.~\cite{WeiNorman,Kolokolov}  Wei and Norman,~\cite{WeiNorman} addressing the problem of a quantum
spin subject to a prescribed classical time-dependent magnetic
field, have elegantly shown that by preforming a {\it non-linear}
transformation from $\bm{\theta}$ to a set of new variables, the time ordered exponent~\eqref{eq_Intro} can be
written as a product of three ordinary exponents (cf. Eq.~\eqref{VarChg}). Even
so, that problem could not be solved in a closed form in general. To find relation between new variables and $\bm{\theta}$ one has to solve Riccati equation. Although the problem of a dynamical magnetic field seems to be even more intricate, in fact, as it was shown by Kolokolov,~\cite{Kolokolov} it is simpler. In this case, one needs to know the Jacobian for the non-linear transformation only. Then, the functional integration over new variables can be performed exactly.

We thus present here exact analytic resuts for the partition function (cf. Eq.~\eqref{GCPF_Gen}) and the tunneling density of states (cf. Eq.~\eqref{GK2}) for the UH with charging and isotropic spin exchange interactions in the presence of Zeeman spliting. We emphasize that our results are valid for arbitrary parameters of the UH. 

In the mesoscopic Stoner regime, near the Stoner instability, $\delta-J\ll \delta$, our general results can be drastically simplified. 
We find that in a wide temperature range $\delta \ll T \ll \delta J/(\delta-J)$ the average zero-field spin susceptibility behaves according to the Curie law with a large effective spin which depends on temperature logarithmically (cf. Eq.~\eqref{chi_fin_f}. The latter is the effect of statistical fluctuations of single-particle levels. A tiny magnetic field $B\sim \sqrt{JT(1-J/\delta)}/g\mu_B$ is enough for the average spin susceptibility to become temperature independent Fermi-liquid like (cf. Eq.~\eqref{SpinSuscRegIIb_Fin}). Here $g$ and $\mu_B$ stand for the $g$-factor and the Bohr magneton, respectively. We find that enhanced spin correlations resulting in a large total spin in the ground state of a QD in the mesoscopic Stoner regime near the Stoner instability, $\delta-J\ll \delta$, can be observed as additional (to Coulomb blockade) non-monotonic behavior in the tunneling density of states (TDOS) at high temperatures $\delta \ll T \ll \delta J/(\delta-J)$. Magnetic field suppresses the spin-related non-monotonic behavior of the TDOS. We mention that some of the results were published in a brief form in Ref.~[\onlinecite{BGK}]. Our main new results concern the effect of Zeeman splitting (cf. Eqs.~\eqref{GK2}, \eqref{FinalNuBB}) and of disorder (cf. Eqs.~\eqref{SpinSuscRegI_Fin}, \eqref{SpinSuscRegIIa_Fin}, \eqref{SpinSuscRegIIb_Fin}, \eqref{chi_fin_f_r3}).

The physics discussed in current work can be tested in QDs made of
materials close to the thermodynamic Stoner instability, e.g.,
Co impurities in a Pd or Pt host, Fe or Mn dissolved in various
transition-metal alloys, Ni impurities in a Pd host, and Co
in Fe grains, as well as new nearly ferromagnetic rare-earth
materials.~\cite{Exp1,Exp2,Exp3} Possibly, the intriguing magnetic behavior 
observed recently in Pd nanoparticles capped with 
different protective systems~\cite{Exp4} is related to the physics of mesoscopic Stoner 
regime.

The outline of the paper is as follows. In Sec.~\ref{Sec.UH_SCS} we introduce the UH, subsequent imaginary time action, and 
partially disentangle charge and spin degrees of freedom in the problem. In Sec.~\ref{WNK_Transform} we introduce the WNK transformation to solve the problem of spin dynamics. In Sections~\ref{Sec.PartFun} and~\ref{Sec.TDOS} with the help of the WNK transformation we derive exact analytic expressions for the grand canonical partition function and for the TDOS corresponding to the UH in the presence of Zeeman splitting and for a given realization of single-particle levels. The analysis presented in Section~\ref{Sec.SpinSusc}) incorporates the effect of  disorder. 
In Sections~\ref{Sec_RegI}-\ref{RegIIISpinSusc} we present a rigorous analysis of
the effect of level fluctuations on the spin susceptibility. The latter is modified by disorder in a strong and significant manner. We refer the less initiate reader to a  semi-qualitative derivation of our results for the average
spin susceptibility (Sec.~\ref{Sec_Chi_Qual}).  In Sec.~\ref{Sec.TDOS.An} we
discuss the dependence of the TDOS on energy, temperature and magnetic field. 
The effect of level fluctuations on the TDOS is semi-qualitatively discussed in Sec.~\ref{Sec_TDOS_Fluc}.
We conclude  the paper with summary of the main results, and brief coments of the amenability of our predictions to experimental tests 
(Sec.~\ref{Sec.Conc}).


\section{Formalism\label{Sec.UH_SCS}}

\subsection{Universal Hamiltonian}

We consider a quantum dot of linear size $L$ in the so-called metallic regime,
whose dimensionless conductance  $g_{\rm Th} = E_{\rm Th}/\delta \gg 1$. Here  $E_{\rm Th}$ is the Thouless energy.
We account for  the following terms of the universal Hamiltonian~\cite{KAA}
\begin{equation}
H =H_0 + H_C+H_S,\qquad H_0= \sum\limits_{\alpha,\sigma} \epsilon_{\alpha,\sigma} a^\dag_{\alpha,\sigma} a_{\alpha,\sigma} . \label{EqUnivHam}
\end{equation}
Here, $\epsilon_{\alpha,\sigma}$ denotes the spin ($\sigma$) dependent single particle  levels. In what follows, we shall assume that the magnetic field $B$ is applied and $\epsilon_{\alpha,\sigma} = \epsilon_\alpha+g\mu_B B \sigma/2$. 

The charging interaction 
\begin{equation}
H_C=E_c \left ( \hat{n} -N_0\right )^2
\end{equation}
accounts for the Coulomb blockade. Here 
\begin{equation}
\hat n \equiv \sum_\alpha \hat n_\alpha=\sum_{\alpha,\sigma} a^\dag_{\alpha,\sigma} a_{\alpha,\sigma}
\end{equation}
  is the particle number operator, and $N_0$ represents the background charge. The term 
\begin{equation}
H_S = -J \bm{S}^2
\end{equation}
represents spin interactions within the dot. Here 
\begin{equation}
\bm{S}=\sum_{\alpha} \bm{s}_\alpha=\frac{1}{2}\sum_{\alpha\sigma\sigma^\prime}  a^\dag_{\alpha,\sigma} \bm{\sigma}_{\sigma\sigma^\prime} a_{\alpha,\sigma^\prime}
\end{equation}
denotes the operator of the total spin of electrons on the dot, with the components of $\bm{\sigma}$ comprising of the Pauli matrices.

We stress that we do not consider the interaction in the Cooper channel in the UH~\eqref{EqUnivHam}. For QDs fabricated 
in 2D electron gas the interaction in the Cooper channel is typically repulsive and, therefore, renormalizes to zero.~\cite{ABG} 
In the absence of spin-orbit interaction the parallel magnetic field does not affect the orbital motion of electrons in a QD (we neglect the effect due to a finite width of 2D electron gas). In this case the statistics of single-particle energies $\epsilon_\alpha$ can be described either by the orthogonal Wigner-Dyson ensemble (class AI) or by the unitary Wigner-Dyson ensemble (class A).~\cite{WD,Zirnbauer} The latter is achieved in a weak perpendicular magnetic field $B_\perp \gtrsim B_{c}  = \Phi_0/(L^2\sqrt{g_{\rm Th}})$ where $\Phi_0$ denotes the flux quantum. In the case of 3D quantum dots realized as small metallic grains, the interaction in the Cooper channel can be attractive, giving rise to superconducting correlations. In this case we assume that there is a weak magnetic field $B \gtrsim B_{c}$ which suppresses the Cooper channel. Therefore, the level statistics is described by the unitary Wigner-Dyson ensemble.

The imaginary time action for the system~\eqref{EqUnivHam} reads
\begin{equation}
S_{\rm tot} = \int_0^\beta \mathcal{L}  d\tau  = \int_0^\beta \Bigl [ \sum\limits_{\alpha}
\overline{\Psi}_{\alpha} (\partial_\tau + \mu)\Psi_{\alpha} - H \Bigr ] d\tau ,
\end{equation}
where $\mu$ is the chemical potential, $\beta=1/T$, and we have introduced the Grassmann variables $\overline{\Psi}_{\alpha} = (\bar\psi_{\alpha\uparrow},\bar\psi_{\alpha\downarrow})^T, \Psi_{\alpha} = (\psi_{\alpha\uparrow},\psi_{\alpha\downarrow})$ to represent electrons on the dot.

Employing the Hubbard-Stratonovich transformation leads to a bosonized form
\begin{eqnarray}
\mathcal{L} &=& \sum_{\alpha}
\overline{\Psi}_{\alpha} \left [ \partial_\tau - \epsilon_\alpha -\frac{g\mu_BB}{2}\sigma_z+\mu  +i \phi+\frac{\bm{\sigma}\cdot \bm{\Phi}}{2} \right ] \Psi_{\alpha}  \notag
\\
&& + \frac{\bm{\Phi}^2}{4J} +\frac{\phi^2}{4E_c}-i N_0 \phi 
\end{eqnarray}
where $\phi$ and $\bm{\Phi}$ are scalar and vector bosonic fields
respectively. The  $SU(2)$ non-Abelian character of the action poses
a serious difficulty. In the presence of the charging interaction only (Abelian $U(1)$ case) the problem can be solved by 
performing a gauge transformation.~\cite{KamenevGefen,EfetovTscherisch,SeldmayrLY} For the case of the charging interaction and spin-exchange interaction of Ising type (Abelian $U(1)\times U(1)$ case) the problem also can be solved by a gauge transformation.~\cite{KG,Boaz}
In the non-Abelian $U(1)\times SU(2)$ case, we start from performing a gauge transformation in the charging sector only.

\subsection{Partial disentanglement of spin and charge}

Our aim is to compute the grand partition function $Z= \Tr \exp (-\beta H + \mu \beta \hat n )$ and Green's function
in the Matsubara time domain
\begin{equation}
G_{\alpha,\sigma_1,\sigma_2}(\tau_1,\tau_2) = - \mathcal{T}_\tau\frac{\Tr a_{\alpha,\sigma_1}(\tau_1) a^\dag_{\alpha,\sigma_2}(\tau_2)  e^{-\beta H + \mu \beta \hat n }}{\Tr e^{-\beta H + \mu \beta \hat n }} .
\end{equation}
Here, $\mathcal{T}_\tau$ denotes a time ordering operation along Matsubara time. In the Lagrangian formalism, the Green's function can be written as
\begin{gather}
G_{\alpha}(\tau_1,\tau_2) = -\frac{1}{Z} \mathcal{T}_\tau \int \mathcal{D}[\overline{\Psi},\Psi,\phi,\bm{\Phi}] \Psi_{\alpha}(\tau_1) \overline{\Psi}_{\alpha}(\tau_2) e^{-S_{\rm tot}} ,\notag \\  
Z = \int\mathcal{D}[\overline{\Psi},\Psi,\phi,\bm{\Phi}]\, e^{-S_{\rm tot}} .
\end{gather}
Let us split the field $\phi(\tau)$ as
\begin{equation}
\phi(\tau) = \tilde{\phi}(\tau) + 2\pi m T + \phi_0, \qquad \int_0^\beta d\tau\, \tilde{\phi}(\tau)=0,
\end{equation}
where the static component of $\phi(\tau)$ obeys inequality $|\phi_0|\leqslant \pi T$. Then the part $\tilde\phi(\tau)+2\pi mT$ of $\phi(\tau)$ can be gauged away (see Refs.~[\onlinecite{KamenevGefen,EfetovTscherisch,KG,SeldmayrLY,Boaz}] for details). The Green's function becomes 
\begin{gather}
{G}_{\alpha}(\tau_{12}) =  \int\limits_{-\pi T}^{\pi T}\frac{d\phi_0}{2\pi T} \,\frac{\mathcal{Z}(\phi_0)}{Z}
{D}(\tau_{12},\phi_0) \, \mathcal{G}_{\alpha}(\tau_{12},\phi_0),
\label{CSSep1} \\
Z = \int_{-\pi T}^{\pi T}\frac{d\phi_0}{2\pi T} \, {D}(0,\phi_0) \mathcal{Z}(\phi_0) , \label{CSSep1Z}
\end{gather}
where $\tau_{12}\equiv\tau_1-\tau_2$. The so-called Coulomb-boson propagator reads
\begin{align}
D(\tau,\phi_0) & =  e^{-E_c|\tau|(1-|\tau|/\beta)} \sum_{k\in \mathbb{Z}} e^{i\phi_0 (\beta k+\tau)}\notag \\
&\times e^{-\beta E_c(k-N_0+\tau/\beta)^2}.
\label{CB}
\end{align}

The Green's function $\mathcal{G}_{\alpha}(\tau_{12},\phi_0) $ on the right hand side of Eq.~\eqref{CSSep1} is defined as 
\begin{gather}
\mathcal{G}_{\alpha}(\tau_{12},\phi_0) = - \mathcal{T}_\tau \int \frac{\mathcal{D}[\overline{\Psi},\Psi,\phi,\bm{\Phi}]}{\mathcal{Z}(\phi_0)} \Psi_{\alpha}(\tau_1) \overline{\Psi}_{\alpha}(\tau_2)  e^{-\mathcal{S}} ,\notag \\  
\mathcal{Z}(\phi_0) = \int\mathcal{D}[\overline{\Psi},\Psi,\phi,\bm{\Phi}]\, e^{-\mathcal{S}} .
\end{gather}
Here the action 
\begin{eqnarray}
\mathcal{S} &=& \int_0^\beta d\tau \sum_{\alpha}
\overline{\Psi}_{\alpha} \left [ \partial_\tau - \epsilon_\alpha +\mu  +i \phi_0+\frac{\bm{\sigma}\cdot \bm{\Phi}}{2} \right ] \Psi_{\alpha}  \notag
\\
&& + \frac{1}{4J}  \int_0^\beta d\tau  \,\bm{\Phi}^2 .
\end{eqnarray}
It can be formally rewritten as  
\begin{equation}
\mathcal{S} =  \int_0^\beta \Bigl [ \sum\limits_{\alpha}
\overline{\Psi}_{\alpha} \partial_\tau \Psi_{\alpha} - \mathcal{H} \Bigr ] d\tau , \quad \mathcal{H} = \mathcal{H}_0 +H_S ,
\end{equation}
where $\mathcal{H}_0$ is given by $H_0$ (Eq.~\eqref{EqUnivHam}) in which $\epsilon_{\alpha,\sigma}$ is replaced by $\tilde{\epsilon}_{\alpha,\sigma}=\epsilon_{\alpha,\sigma} -\mu+i \phi_0$:
\begin{equation}
\mathcal{H}_0= \sum\limits_{\alpha,\sigma} \tilde{\epsilon}_{\alpha,\sigma} a^\dag_{\alpha,\sigma} a_{\alpha,\sigma} . \label{H0_1}
\end{equation}
Remarkably, the charge and spin degrees of freedom are almost disentangled in the action $\mathcal{S}$. The latter involves only 
the exchange interaction $H_S$. The remnant traces of the charging interaction $H_C$ are encoded in the variable $\phi_0$, leading to a small imaginary shift of the chemical potential.

\section{Wei-Norman-Kolokolov transformation\label{WNK_Transform}}

The evaluation of the Green's function $\mathcal{G}_\alpha(\tau_{12})$ is more convenient to perform in the Hamiltonian formalism.
Then it can be written as 
\begin{gather}
\mathcal{G}_{\alpha\sigma_1\sigma_2}(\tau) = \frac{1}{\mathcal{Z}}
\begin{cases} 
- \mathcal{K}_{\alpha\sigma_1\sigma_2}(-i\tau,-i\tau+i\beta), & \, \tau>0 ,\\
\mathcal{K}_{\alpha\sigma_1\sigma_2}(-i\tau-i\beta,-i\tau), & \, \tau\leqslant 0  .
\end{cases}
\end{gather}
Here $\mathcal{Z} = \Tr \exp(-\beta \mathcal{H})$ and we introduce
\begin{equation}
\mathcal{K}_{\alpha\sigma_1\sigma_2}(t_+,t_-) = \Tr e^{-i t_+ \mathcal{H}} a^\dag_{\alpha,\sigma_2}  e^{i t_- \mathcal{H}}  a_{\alpha,\sigma_1} .
\label{Eq11}
\end{equation}
Next, using the following set of transformations for the evolution operator (we recall that $\mathcal{H}_0$ and $H_S$ commute), we write
\begin{eqnarray}
e^{\mp i t J\bm{S}^2} &=&  \lim\limits_{N\to\infty}\prod_{n=1}^N e^{\mp i t J\bm{S}^2/N} \notag \\
&=& \lim\limits_{N\to\infty} \prod_{n=1}^N \int d \bm{\theta}_n\, e^{\pm \frac{i}{4J} t \bm{\theta}^2_n/N} \prod_\alpha e^{i t \bm{\theta}_n \bm{s}_\alpha/N} \notag \\
&=&   \int \mathcal{D}[\bm{\theta}]\, e^{\pm \frac{i}{4J} \int_0^t dt^\prime\, \bm{\theta}^2} \prod_\alpha \mathcal{T} e^{i \int_0^t dt^\prime\, \bm{\theta} \bm{s}_\alpha} , \label{TimeOrd}
\end{eqnarray}
and obtain
\begin{gather}
\mathcal{K}_{\alpha\sigma_1\sigma_2}(t_+,t_-) = \prod_{p=\pm} \int \mathcal{D}
[\bm{\theta}_p]e^{- \frac{i p}{4J} \int_0^{t_p} dt^\prime\, \bm{\theta}_p^2}  \Tr \Bigl [ e^{-i t_+ \mathcal{H}_0} \notag \\
 \times \prod_\gamma
\mathcal{A}_\gamma^{(+)}
a^\dag_{\alpha,\sigma_2}  e^{i t_- \mathcal{H}_0}
\prod_\eta
\mathcal{A}_\eta^{(-)}  a_{\alpha,\sigma_1}\Bigr ]. \label{K1}
\end{gather}
Here we have introduced the bosonic fields $\bm{\theta}_p$, $p=\pm$, and 
\begin{equation}
\label{eq_A}
 \mathcal{A}_\alpha^{(p)} = \mathcal{T} \exp \left ( i \int_0^{t_p}
dt^\prime\, \bm{\theta}_p \bm{s}_\alpha\right ) .
\end{equation}

In Eq.~\eqref{TimeOrd} we have missed the correct normalization factor (depending on $J$) because in the Hubbard-Stratonovich decoupling we were not pedantic enough concerning the normalization factor. For the computation of the Green's function it is irrelevant due to the cancelation of normalization factors. For the partition function we restore it later on by comparison with known limiting cases. Note that while $\mathcal{H}$ is time independent,
the factors $\mathcal{A}_\alpha^{(p)}$ involve time ordering
($\mathcal{T}$). This is due to the non-commutativity of the
spin-operators $\bm{s}_\alpha$.

In order to overcome the intricacy of time-ordering we apply the WNK transformation~\cite{WeiNorman,Kolokolov}
of variables in the functional integral~\eqref{K1} (see Appendix~\ref{Appendix_WNK} for details):
\begin{gather}
\theta_{p}^z = \rho_{p} - 2 \kappa_{p}^p\kappa_{p}^{-p},\, \frac{\theta_{p}^x- ip \theta_{p}^y}{2}=\kappa_{p}^{-p},\notag \\
\frac{\theta_{p}^x+i p \theta_{p}^y}{2}=- i p \dot{\kappa}_{p}^p +\rho_{p} \kappa_{p}^p - (\kappa_{p}^p)^2
\kappa_{p}^{-p} . \label{VarChg}
\end{gather} 
Here new variables $\rho_+, \kappa_+^+, \kappa_+^-$ correspond to $\bm{\theta}_+$ whereas 
new variables $\rho_-, \kappa_-^-, \kappa_-^+$ are introduced instead of $\bm{\theta}_-$. 

The WNK transformation recasts the time-ordered exponent as a product of simple Abelian ones:
\begin{align}
\mathcal{A}_\gamma^{(p)}  &=  \exp \Bigl [p \hat s_\gamma^{-p} \kappa_{p}^p (t_{p})\Bigr] \exp \left [ i \hat s_\gamma^z\int_0^{t_{p}}
dt^\prime \rho_{p}(t^\prime)\right ] \notag \\
&\times \exp \left [ i \hat s_\gamma^p \int_0^{t_{p}} dt^\prime \kappa_{p}^{-p}(t^\prime) e^{- ip \int_0^{t^\prime} d\tau \rho_{p}(\tau)} dt^\prime \right ]
.\label{A2P}
\end{align}
Here $s_\gamma^\pm = s_\gamma^x\pm i
s_\gamma^y$, and we employ the initial condition $\kappa_{p}^p(0)
=0$ (the origin of this initial condition is discussed in Appendix~\ref{Appendix_WNK}). We stress that Eqs~\eqref{VarChg} and \eqref{A2P} are
valid for a general spin operator.
Originally, the field variables
$\bm{\theta}_{p}$ were real, but before the change of
variables~\eqref{VarChg} we have rotated the contour of integration
in the complex plane. This procedure does not interfere with convergence of the Gaussian integrals.
In order to preserve the number of field variables (three) we impose
the following constraints on the otherwise arbitrary new complex
variables: $\rho_{p}=-\rho_{p}^*$ and $\kappa_{p}^+ = (\kappa_{p}^-)^* $. 
The Jacobian of the Wei-Norman-Kolokolov transformation~\eqref{VarChg}
is given as (see Appendix~\ref{Appendix_WNK})
\begin{equation}
\mathcal{J} = \prod_{p=\pm} \exp \left (  \frac{i p}{2} \int_0^{t_p} dt\,
\rho_p(t) \right ). \label{Jac_WNK}
\end{equation}
In terms of new variables the quantity $\mathcal{K}_{\alpha\sigma_1\sigma_2}(t_+,t_-)$ can be then rewritten as
\begin{gather}
\mathcal{K}_{\alpha\sigma_1\sigma_2}(t_+,t_-)  = \prod_{p=\pm}
\int \mathcal{D}[\rho_{p}, \kappa_{p}^{\pm p}]  e^{\frac{p}{4iJ}\! \int_0^{t_p}\! dt(\rho_p^2-4 i p\dot{\kappa}_p^p \kappa_p^{-p})}
 \notag 
\\
 \times e^{\frac{i p}{2} \int_0^{t_p} dt \rho_p(t)} \mathcal{C}_{\alpha\sigma_1\sigma_2}(t_+,t_-)
\prod_{\gamma\neq \alpha} \mathcal{B}_\gamma(t_+,t_-) ,\label{K2}
\end{gather}
with $\mathcal{C}_{\alpha\sigma_1\sigma_2}$ and $\mathcal{B}_\gamma$ given in terms of single-particle traces:
\begin{eqnarray}
\mathcal{C}_{\alpha\sigma_1\sigma_2} &=& 
\tr \Bigl [ e^{-i \hat h_\alpha   t_+} \mathcal{A}_\alpha^{(+)}(t_+) a^\dag_{\alpha,\sigma_2}e^{i \hat h_\alpha  t_-} \mathcal{A}_\alpha^{(-)}(t_-) a_{\alpha,\sigma_1} \Bigr ] , \notag \\
\mathcal{B}_\gamma &=& 
\tr \Bigl  [e^{-i  \hat h_\gamma t_+} \mathcal{A}_\gamma^{(+)}(t_+)e^{i  \hat h_\gamma t_-} \mathcal{A}_\gamma^{(-)}(t_-) \Bigr ] . 
\end{eqnarray}
Here $\hat h_\alpha = \sum_\sigma \tilde{\epsilon}_{\alpha,\sigma}\hat{n}_{\alpha,\sigma}$.
The expression for $\mathcal{Z}$ can be obtained from Eq.~\eqref{K2} by the
substitution of $\mathcal{B}_\alpha$ for $\mathcal{C}_{\alpha\sigma_1\sigma_2}$: 
\begin{gather}
\mathcal{Z} = \prod_{p=\pm}
\int \mathcal{D}[\rho_{p}, \kappa_{p}^{\pm p}] e^{\frac{p }{4iJ} \int_0^{t_p}dt (\rho_p^2-4i p \dot{\kappa}_p^p \kappa_p^{-p})}
 \notag 
\\
\times e^{\frac{i p}{2} \int_0^{t_p} dt \rho_p(t)} \prod_{\gamma} \mathcal{B}_\gamma(t_+,t_-) . \label{ZK2}
\end{gather}

Simplifying expression~\eqref{A2P} with the help of identity $(\hat s_\gamma^p)^2=0$ (valid for spin $1/2$) and evaluating the single-particle traces, we find the following result:
\begin{eqnarray}
\mathcal{B}_\gamma &=& 1+e^{-2i\tilde\epsilon_\gamma (t_+-t_-) } \notag \\
&+&2 e^{- i\tilde\epsilon_\gamma (t_+-t_-)}  \cos\Bigl [ \frac{1}{2}  \sum_{p=\pm}\int\limits_0^{t_p} dt  \tilde{\rho}_p(t) \Bigr ] \notag \\
&+& \prod_{p=\pm} e^{-i p \tilde\epsilon_\gamma t_p} \exp \Bigl [ \frac{i p}{2} \int\limits_0^{t_p}dt \tilde{\rho}_p(t)\Bigr ] \notag\\
&\times & \left [ p\tilde{\kappa}_p^p(t_p)+i \int\limits_0^{t_{-p}}dt\,  \tilde{\kappa}_{-p}^p(t) e^{i p \int\limits_0^t dt^\prime \tilde{\rho}_{-p}(t^\prime)}\right ] .
\label{BA1}
\end{eqnarray}
Here the presence of Zeeman splitting is taken into account by means of the variables ($b=g\mu_B B$)
\begin{equation}
\tilde{\rho}_p(t) = \rho_p(t) - p b,\,\qquad  \tilde{\kappa}^p_{\pm p}(t)= \kappa^p_{\pm p}(t) e^{\pm ib t} .
\end{equation}
The evaluation of the single-particle traces yields the following non-trivial matrix structure of $\mathcal{C}_{\alpha}$ in the spin space:
\begin{align}
\mathcal{C}_{\alpha\uparrow\uparrow} &= e^{-2 i \tilde\epsilon_\alpha t_+} \sum_{p=\pm} e^{i\tilde\epsilon_\alpha t_{p}}
e^{\frac{i p }{2}\int_0^{t_p}dt\tilde{\rho}_p(t)} , \notag\\
\mathcal{C}_{\alpha\uparrow\downarrow} &=e^{-2i\tilde\epsilon_\alpha t_+} \Bigl [ i e^{i\tilde\epsilon_\alpha t_+} e^{\frac{i}{2}\int_0^{t_+}dt \tilde\rho_+(t)}
\int_0^{t_+} dt^\prime \tilde{\kappa}_+^-(t^\prime) \notag \\
&\times e^{-i\int_0^{t^\prime}d\tau \tilde{\rho}_+(\tau)}
+ e^{i\tilde\epsilon_\alpha t_-} \tilde{\kappa}_-^-(t^\prime) e^{-\frac{i}{2}\int_0^{t_-}dt \tilde\rho_-(t)}
\Bigr ] ,\notag \\
\mathcal{C}_{\alpha\downarrow\uparrow} &=e^{-2i\tilde\epsilon_\alpha t_+} \Bigl [ -i e^{i\tilde\epsilon_\alpha t_-} e^{-\frac{i}{2}\int_0^{t_-}dt \tilde\rho_-(t)}
\int_0^{t_-} dt^\prime \tilde{\kappa}_-^+(t^\prime) \notag \\
&\times e^{i\int_0^{t^\prime}d\tau \tilde{\rho}_-(\tau)}
+ e^{i\tilde\epsilon_\alpha t_+} \tilde{\kappa}_+^+(t^\prime) e^{\frac{i}{2}\int_0^{t_+}dt \tilde\rho_+(t)}
\Bigr ] ,\notag \\
\mathcal{C}_{\alpha\downarrow\downarrow} &=e^{-2 i \tilde\epsilon_\alpha t_+} \sum_{p=\pm} e^{i\tilde\epsilon_\alpha t_{p}}
e^{-\frac{i p }{2}\int_0^{t_p}dt\tilde{\rho}_p(t)} \Bigl [ 
1 +i p \tilde{\kappa}_p^p(t_p)\notag\\
&\times e^{i p \int_0^{t_p}dt \tilde\rho_p(t)}
\int_0^{t_p} dt^\prime \tilde{\kappa}_p^{-p}(t^\prime) e^{-ip\int_0^{t^\prime}d\tau \tilde{\rho}_p(\tau)}
\Bigr ] .
\label{CA1}
\end{align}
We emphasize that the WNK transformation explicitly breaks the symmetry $S_z\to -S_z$ together with $b\to -b$. For example, $\mathcal{C}_{\alpha\downarrow\downarrow}$ cannot be obtained from $\mathcal{C}_{\alpha\uparrow\uparrow}$ by reversing sign of the magnetic field $b$. We shall see below how the symmetry restores.

\section{Exact expression for the partition function\label{Sec.PartFun}}

The partition function $\mathcal{Z}$ is given by Eq.~\eqref{ZK2}. We start from integration over the fields $\kappa_p^p$ and $\kappa_{-p}^p$. 
The expression~\eqref{BA1} for  $\mathcal{B}_\gamma$ is bilinear form of the fields $\kappa_p^p$ and $\kappa_{-p}^p$. By using the following identity
\begin{equation}
x = - \oint\limits_{|z|=1} \frac{dz}{2\pi i}\frac{e^{-z x}}{z^2},
\end{equation}
we rewrite Eq.~\eqref{ZK2} as
\begin{align}
\mathcal{Z} &= \prod_{p=\pm}
\int \mathcal{D}[\rho_{p}, \kappa_{p}^{\pm p}] e^{-\frac{i p }{4J} \int_0^{t_p}dt (\rho_p^2-4i p \dot{\kappa}_p^p \kappa_p^{-p} -2J \rho_p(t))} 
 \notag 
\\
&\times \left ( \prod_{\gamma}  \oint\limits_{|z_\gamma|=1} \frac{i dz_\gamma}{2\pi z_\gamma^2}\right )\exp \left ( -\sum\limits_\gamma z_\gamma \mathcal{B}_\gamma(t_+,t_-)\right ) . \label{ZK3}
\end{align}
Then the functional integral over the fields $\kappa_p^{\pm p}$ becomes Gaussian. As shown in Appendix~\ref{Appendix_IntWNK}, due to the specific form of the initial conditions, they have simple dynamics and can be integrated out exactly. The result is
\begin{align}
\mathcal{Z}&=J^2 
 \left ( \prod_{p=\pm}
\int \mathcal{D}[\rho_{p}] e^{-\frac{i p }{4J} \int_0^{t_p}dt \rho_p^2+\frac{i p}{2} \int_0^{t_p} dt \rho_p(t)} \right )
 \notag 
\\
&\times \left ( \prod_{\gamma}  \oint\limits_{|z_\gamma|=1} \frac{i dz_\gamma}{2\pi z_\gamma^2}\right ) 
e^{-w-2 v \cos\Bigl [ \frac{1}{2}  \sum\limits_{p=\pm}\int\limits_0^{t_p} dt  \tilde{\rho}_p(t) \Bigr ] }
\notag \\
&\times \int_0^\infty dy \, 
\exp\Biggl [ -y- i J y \left ( \prod_{p=\pm} e^{\frac{ip}{2} \int_0^{t_p}dt \rho_p} \right )\notag \\
&\times  \left ( v \sum\limits_{p=\pm} p e^{\frac{i p b}{2}(t_+-t_-)}
\int_0^{t_p} dt\, e^{-i p \int_0^t dt^\prime \rho_p(t^\prime)} \right ) \Biggr ] , \label{ZK4}
\end{align}
where
\begin{align}
w &=\sum_{\gamma} z_\gamma \left [ 1+e^{- 2i \tilde\epsilon_\gamma (t_+-t_-)} \right ], \notag \\
 v &=
\sum_{\gamma} z_\gamma e^{- i\tilde\epsilon_\gamma (t_+-t_-)}.
\end{align}

To transform Eq.~\eqref{ZK4} to more standard form, let us introduce new variables
\begin{equation}
\xi_p(t) = i p \int_0^t dt^\prime \rho_p(t^\prime) + \xi_p(0) .
\end{equation}
Here $\xi_p(0)$ is an arbitrary constant. Then the partition function $\mathcal{Z}$ can be written as
\begin{align}
\mathcal{Z}= & J^2
 \left ( \prod_{\gamma}  \oint\limits_{|z_\gamma|=1} \frac{i dz_\gamma}{2\pi z_\gamma^2}\right )  \int_0^\infty dy \, e^{-y-w}\Biggl [  \prod_{p=\pm} \int \mathcal{D}[\xi_{p}] \notag \\
 \times &
\exp \left ( i p \int_0^{t_p}dt\, {\mathcal{L}}_p + \frac{\xi_p(t_p)-\xi_p(0)}{2}\right ) \Biggr ]\notag \\
\times & \exp \left \{ -2 v \cosh \Biggl  ( \sum\limits_{p=\pm} \frac{p}{2} [\xi_p(t_p)-\xi_p(0)- i b t_p] \Biggr ) \right \} .
 \label{ZK5}
\end{align}
The functional integral over fields $\xi_p$ in Eq.~\eqref{ZK5} is of the Feynman-Kac type. The quantity
\begin{equation}
{\mathcal{L}}_p = \frac{1}{4J} \dot{\xi}_p^2 - J y v \left ( \prod_{q=\pm} e^{\frac{ \xi_q(t_q) + p q \xi_q(0) + i p b q t_q}{2} } \right ) e^{-\xi_p}  
\end{equation}
plays a role of Lagrangian. It is convenient to perform a shift of variables $\xi_p$ and introduce new variables
\begin{equation}
\tilde{\xi}_p(t) = \xi_p(t) - \frac{1}{2} \sum_{q=\pm} \left [ \xi_q(t_q) + p q \xi_q(0) + i p b q t_q \right ]  .
\end{equation}
Then the expression for the partition function $\mathcal{Z}$ acquires exactly the same form as given in Eq.~\eqref{ZK5} with the following substitutions:
$\xi_p(t_p) \to \tilde{\xi}_p(t_p)$, $\xi_p(0) \to \tilde{\xi}_p(0)$, ${\mathcal{L}}_p \to \tilde{\mathcal{L}_p}$,
where
\begin{equation}
\tilde{\mathcal{L}_p} = \frac{1}{4J} \dot{\tilde\xi}_p^2 - \frac{J}{4} e^{-\tilde\xi_p} .\label{LK1}
\end{equation}
We mention that the new variables $\tilde{\xi}_p$ are, in fact, independent of the values of $\xi_p(0)$. They obey the following constraints
\begin{eqnarray}
\sum_{p=\pm} \tilde{\xi}_p(t_p)+2 \ln 4 y v &=& 0, \\
\sum_{p=\pm} p [ \tilde{\xi}_p(0) +i b t_p ] &=& 0 .
\end{eqnarray}
In what follows we shall omit the tilde signs on the variables $\tilde\xi_p$. At this point it is convenient to express Eq.~\eqref{ZK5} in terms of matrix elements for the one-dimensional quantum mechanics with the Hamiltonian
\begin{equation}
 H_J = - J \frac{\partial^2}{\partial \xi^2}
+ \frac{J}{4} e^{-\xi} .
\end{equation}
Then the partition function $\mathcal{Z}$ becomes
\begin{align}
\mathcal{Z} &= 
 \left ( \prod_{\gamma}  \oint\limits_{|z_\gamma|=1} \frac{idz_\gamma}{2\pi z_\gamma^2}\right ) \int_0^\infty \frac{J^2dy}{4y v} \, e^{-y-w} \Biggl (\prod_{p=\pm} \int d\xi_p d\xi_p^\prime \,\notag \\
 &\times  e^{-\xi_p^\prime/2}  \Biggr ) \delta\left (\sum_{p=\pm}\xi_p+2\ln 4y v\right ) \delta\left (\sum_{p=\pm} p [ \xi_p^\prime +i b t_p] \right ) 
 \notag \\
& \times 
 e^{-2v \cosh[(\xi_+-\xi_-)/2]}
 \langle \xi_+ | e^{-i H_J t_+}|\xi_+^\prime \rangle
  \langle \xi_-^\prime | e^{i H_J t_-}|\xi_- \rangle .
  \label{ZK6}
\end{align}
The Hamiltonian $H_J$ of the one-dimensional quantum mechanics is exactly solvable. Its eigenfunctions are spanned by modified Bessel functions $K_{2i\nu}$:
\begin{equation}
\langle \nu | \xi\rangle = \frac{2}{\pi}\sqrt{\nu \sinh(2\pi\nu)} K_{2i\nu}(e^{-\xi/2})  ,
\end{equation}
where $\nu$ is a real parameter. The corresponding eigenvalues of $H_J$ are equal to $J\nu^2$: $H_J|\nu\rangle = J\nu^2|\nu\rangle$. 

Next we perform integration over $y$ in Eq.~\eqref{ZK6}. Then with the help of the following identity (see formula 6.794.11 on page 794 of Ref.~[\onlinecite{GR}])
\begin{gather}
 \int_0^\infty d\nu\, \nu \sinh(2\pi\nu) K_{2i\nu}(e^{-\xi_+/2})K_{2i\nu}(e^{-\xi_-/2})K_{2i\nu}(2v) \notag \\
 = \frac{\pi^2}{16}
\exp \left ( -\frac{1}{4v}e^{-\frac{\xi_++\xi_-}{2}}-2 v \cosh \frac{\xi_+-\xi_-}{2}\right ) ,\label{Iden1}
\end{gather}
we integrate over $\xi_p$ and obtain
\begin{align}
\mathcal{Z} &= \frac{2J^2}{\pi^2}
 \left ( \prod_{\gamma}  \oint\limits_{|z_\gamma|=1} \frac{idz_\gamma}{2\pi z_\gamma^2}\right ) \frac{e^{-w}}{v}  \int_0^\infty d\nu \,\nu \sinh(2\pi \nu) \notag \\
& \times K_{2i\nu}(2v) 
 \left (\prod_{p=\pm} \int d\xi_p^\prime \, e^{-\xi_p^\prime/2}  K_{2i\nu}(e^{-\xi_p^\prime/2})\right ) \notag \\
&  \times \delta\left (\sum_{p=\pm} p [ \xi_p^\prime +i b t_p] \right )e^{-iJ\nu^2(t_+-t_-)}  . 
 \label{ZK7}
\end{align}
Next, using the identity (see formula 6.521.3 on page 658 of Ref.~[\onlinecite{GR}])
\begin{equation}
\int_0^\infty dx\, x K_\nu(ax) K_\nu(bx) =  \frac{\pi (ab)^{-\nu} (a^{2\nu}-b^{2\nu})}{2\sin(\pi\nu)(a^2-b^2)} ,
\end{equation}
we perform integration over $\xi_+^\prime$ and $\xi_-^\prime$. With the help
of the well-known integral representation of the modified Bessel function
\begin{equation}
K_{2i\nu}(2d) = \frac{1}{2} \int_{-\infty}^\infty dh\, e^{-2 d \cosh h+ 2i\nu h} ,
\end{equation}
we integrate over the variable $\nu$. Finally, integration over $z_\gamma$ can be performed, and
we find
\begin{eqnarray}
\mathcal{Z}&=& \frac{\sqrt{J^3}}{2\sqrt{\pi\beta}} e^{-\beta b^2/4J}
\int_{-\infty}^\infty dh\, \sinh(h) \frac{\sinh(b h/J)}{\sinh(\beta b/2)}\notag \\
&\times & e^{- h^2/\beta J}  \prod_{\gamma,\sigma}  \left (1+e^{-\beta \tilde \epsilon_\gamma-h\sigma} \right ) .
\end{eqnarray}

During the set of transformations we omitted normalization factors which depend on the parameter $J$. In order to restore them, one can compute the 
partition function $\mathcal{Z}$ for a single and two-level cases. Then  one finds that the following transformation is necessary
\begin{equation}
\mathcal{Z} \to \frac{2}{J^2} e^{-\beta J/4} \mathcal{Z}. \label{NormFact}
\end{equation}
Hence, we obtain the following result for the partition function corresponding to the Hamiltonian $\mathcal{H}$:
\begin{eqnarray}
\mathcal{Z} &=& \frac{1}{\sqrt{\pi \beta J}} e^{-\beta (b^2+J^2)/4J}
\int_{-\infty}^\infty dh\, \sinh(h) \frac{\sinh(b h/J)}{\sinh(\beta b/2)}\notag \\
&\times & e^{-h^2/\beta J} \prod_{\gamma,\sigma}  \left (1+e^{-\beta \tilde \epsilon_\gamma-h\sigma} \right ) 
. \label{ZphiFinal_B}
\end{eqnarray}

With the help of Eq.~\eqref{CSSep1Z}, the grand canonical partition function for the full Hamiltonian~\eqref{EqUnivHam} can be written as
\begin{eqnarray}
Z &=&\frac{\sqrt{\beta}}{\sqrt{\pi J}} e^{-\beta (b^2+J^2)/4J}
 \sum_{n\in \mathbb{Z}} e^{-\beta E_c(n-N_0)^2} \int_{-\pi/\beta}^{\pi/\beta} \frac{d\phi_0}{2\pi} \notag \\
 &\times &   e^{i\beta \phi_0 n} \int_{-\infty}^\infty dh\, \sinh(h) \frac{\sinh(b h/J)}{\sinh(\beta b/2)}  e^{-h^2/\beta J}\notag \\
&\times &  \prod_\sigma e^{-\beta \Omega_0(\mu-i\phi_0+h\sigma/\beta)}
. \label{ZphiFinal_BFull}
\end{eqnarray}
To integrate over the variables $\phi_0$ and $h$ we can use the following identity for the grand partition function of free electrons
\begin{equation}
e^{-\beta \Omega_0(\mu)}=\prod_{\gamma}  \left (1+e^{-\beta (\epsilon_\gamma-\mu)} \right ) = \sum_{N=0}^\infty Z_N e^{\beta \mu N} ,
\end{equation}
where the canonical partition function of $N$ noninteracting spinless electrons is given by Darwin-Fowler integral:
\begin{equation}
Z_N \equiv\int_0^{2\pi} \frac{d\theta}{2\pi} e^{-i\theta N}
\prod_{\gamma} \left( 1+e^{i\theta-\beta\epsilon_\gamma}\right)  .
\end{equation}
 Hence we find another representation of the grand canonical partition function for the Hamiltonian~\eqref{EqUnivHam}:
\begin{align}
Z & =\sum_{n_{\uparrow},n_{\downarrow}\in \mathbb{Z}} \frac{\sinh\frac{\beta b (2m+1)}{2}}{\sinh\frac{\beta b}{2}}
Z_{\nup} Z_{\ndn} e^{-\beta E_c(n-N_0)^2+\beta\mu n} \notag \\
& \hspace{2cm} \times  e^{\beta Jm(m+1)]} .
\label{GCPF_Gen}
\end{align}
Here $\nup(\ndn)$ represents the number of spin-up (spin-down)
electrons, the total number of electrons  $n=\nup+\ndn$, and
$m=(\nup-\ndn)/2$. Note that for $m \geqslant  0$ ($m<0$) the total
spin  $S=m$ ($S=-m-1$), respectively. Different terms in Eq.~\eqref{GCPF_Gen} have clear physical meaning. The quantity $E_c(n-N_0)^2- Jm(m+1)$ is the interaction energy of the state with $n_\uparrow$ and $n_\downarrow$ electrons. The factors $Z_{n_\uparrow}$ and $Z_{n_\downarrow}$ 
take into account the contributions from the single-particle energies. The $b$-dependent factor 
\begin{equation}
\sinh\Bigl [\frac{\beta b (2m+1)}{2}\Bigr ]\Bigl /\sinh\Bigl [\frac{\beta b}{2}\Bigr ] \equiv \sum_{S_z=-m}^m \exp(\beta b S_z) 
\end{equation}
represents the partition function for spin $S=m$ in the presence of Zeeman splitting.
Finally, we mention that Eq.~\eqref{GCPF_Gen} coincides with the result obtained in Refs.~[\onlinecite{AlhassidRupp,AlhassidHernando}] by other approach.


\section{Exact expression for the tunneling density of states\label{Sec.TDOS}}

To derive an expression for the TDOS we begin from evaluation of 
the correlation function $\mathcal{K}_{\alpha\sigma_1\sigma_2}$ which is given by Eq.~\eqref{K2}. 
As in the previous section, we start from integration over the fields $\kappa_p^{\pm p}$. The quantities $\mathcal{C}_{\alpha\uparrow\downarrow}$ and 
$\mathcal{C}_{\alpha\downarrow\uparrow}$ are first order in $\kappa_p^{\pm p}$. Therefore, the correlation functions 
$\mathcal{K}_{\alpha\uparrow\downarrow}$ and $\mathcal{K}_{\alpha\downarrow\uparrow}$ vanish after integration over the fields $\kappa_p^{\pm p}$. There is difference between $\mathcal{C}_{\alpha\uparrow\uparrow}$ and $\mathcal{C}_{\alpha\downarrow\downarrow}$. The former is independent of the fields $\kappa_p^{\pm p}$ whereas the latter does. Such an asymmetry is due to our choice in parameterization of the time ordered exponents (see Eq.~\eqref{A2P}). In what follows, we shall evaluate the correlation function $\mathcal{K}_{\alpha\uparrow\uparrow}$. 
Then integration over the fields $\kappa_p^{\pm p}$ can be done in exactly the same way as in the previous section for $\mathcal{Z}$,
since the quantity $\mathcal{C}_{\alpha\uparrow\uparrow}$ is independent of the variables $\kappa_p^{\pm p}$. We thus obtain 
\begin{align}
\mathcal{K}_{\alpha\uparrow\uparrow} &= J^2
 \left ( \prod_{p=\pm}
\int \mathcal{D}[\rho_{p}] e^{-\frac{i p }{4J} \int_0^{t_p}dt \rho_p^2+\frac{i p}{2} \int_0^{t_p} dt \rho_p(t)} \right )\notag \\
& \times 
 \left ( \prod_{\gamma\neq \alpha}  \oint\limits_{|z_\gamma|=1} \frac{idz_\gamma}{2\pi z_\gamma^2}\right ) e^{-w_\alpha-2 v_\alpha \cos\Bigl [ \frac{1}{2}  \sum\limits_{p=\pm}\int\limits_0^{t_p} dt  \tilde{\rho}_p(t) \Bigr ] }\notag \\
\notag \\
&\times   
 \int_0^\infty dy \, e^{-y}
\exp\Biggl [ - i J y \left ( \prod_{p=\pm} e^{\frac{ip}{2} \int_0^{t_p}dt \rho_p} \right ) \notag \\
& \hspace{.5cm}\times \left ( v \sum\limits_{p=\pm} p e^{\frac{i p b}{2}(t_+-t_-)}
\int_0^{t_p} dt\, e^{-i p \int_0^t dt^\prime \rho_p(t^\prime)} \right ) \Biggr ] \notag \\
&\times e^{-2 i \tilde\epsilon_\alpha t_+} \sum_{p=\pm} e^{i\tilde\epsilon_\alpha t_{p}}
e^{\frac{i p }{2}\int_0^{t_p}dt\tilde{\rho}_p(t)}
, \label{K4}
\end{align}
where
\begin{align}
w_\alpha & =\sum_{\gamma\neq\alpha} z_\gamma (1+e^{- 2i \tilde\epsilon_\gamma (t_+-t_-)}), \notag \\
v_\alpha &=
\sum_{\gamma\neq\alpha} z_\gamma e^{- i\tilde\epsilon_\gamma (t_+-t_-)}.
\end{align}

As in the previous section we perform a transformation of variables from $\rho_p$ to $\tilde{\xi}_p$ and  write the result in the Hamiltonian formalism (omitting tilde signs):
\begin{align}
\mathcal{K}_{\alpha\uparrow\uparrow} &=  
 \left ( \prod_{\gamma\neq\alpha}  \oint\limits_{|z_\gamma|=1} \!\frac{idz_\gamma}{2\pi z_\gamma^2}\right )  \int_0^\infty \frac{J^2 dy}{4y v_\alpha} \, e^{-y-w_\alpha-2 i \tilde\epsilon_\alpha t_+}  
 \notag \\
 &\hspace{-0.5cm}\times \Biggl (\prod_{p=\pm} \int  d\xi_p d\xi_p^\prime  e^{-\xi_p^\prime/2}  \Biggr ) \Biggl ( \sum_{p=\pm} e^{i\tilde\epsilon_\alpha t_{p}} e^{-\frac{i b t_p}{2}}
e^{\frac{\xi_p-\xi_p^\prime}{2}} \Biggr )
 \notag \\
&\hspace{-0.5cm}\ \times \delta\left (\sum_{p=\pm}\xi_p+2\ln 4y v_\alpha\right ) \delta\left (\sum_{p=\pm} p [ \xi_p^\prime +i b t_p] \right ) 
 \notag \\
&\hspace{-0.5cm}\ \times e^{-2v_\alpha\cosh[(\xi_+-\xi_-)/2]} \langle \xi_+ | e^{-i H_J t_+}|\xi_+^\prime \rangle
  \langle \xi_-^\prime | e^{i H_J t_-}|\xi_- \rangle .
    \label{K6}
\end{align}
Next we perform integration over $y$ in Eq.~\eqref{K6}. With the help of the identity~\eqref{Iden1} we obtain the following result
\begin{align}
\mathcal{K}_{\alpha\uparrow\uparrow}& = e^{-2 i \tilde\epsilon_\alpha t_+}  
 \left ( \prod_{\gamma\neq\alpha}  \oint\limits_{|z_\gamma|=1} \frac{i dz_\gamma}{2\pi z_\gamma^2}\right ) \frac{J^2}{2v_\alpha} e^{-w_\alpha}  
\int_0^\infty d\nu \notag \\
& \times K_{2i\nu}(2v_\alpha) \sum_{p=\pm}\Biggl [ e^{i\tilde\epsilon_{\alpha\downarrow} t_{p}}  e^{i p J \nu^2 t_{-p}} 
 \int d\nu_1 e^{-i p J\nu_1^2 t_p}  \notag \\
& \times \langle \nu | e^{\xi/2} |\nu_1\rangle Q_{\nu\nu_1}\left (e^{\frac{ip b(t_+-t_-)}{4}}\right )\Biggr ] ,
\label{K7}
\end{align}
where 
\begin{align}
Q_{\nu\nu_1}(z) &= \frac{8 z}{\pi^2} \left [ \nu\nu_1 \sinh(2\pi\nu)\sinh(2\pi\nu_1)\right ]^{1/2}  \notag \\
& \times 
\int_0^\infty d\eta\, \eta^2 K_{2i\nu}(\eta/z)K_{2i\nu_1}(z \eta) .\label{Qdef}
\end{align}
Using the identity (see formula 6.576.4 on page 676 of Ref.~[\onlinecite{GR}]), we find
\begin{align}
Q_{\nu\nu_1}(z) &= \frac{1}{2} z^{-2 -4i\nu} \left [ \prod_{\sigma,\sigma^\prime=\pm}\Gamma\left (\frac{3}{2}+ i\sigma\nu+i\sigma^\prime\nu_1\right )\right ]  \notag \\
& \times {}_2F_1\left (\frac32+i\nu+i\nu_1,\frac32+i\nu-i\nu_1,3;1-z^{-4}\right ) ,
\label{QIden}
\end{align}
where $\Gamma(x)$ and ${}_2F_1(a,b,c;x)$ stand for the Gamma and hypergeometric functions, respectively. With the help of the following relation between the modified Bessel functions 
\begin{equation}
\frac{1}{\eta} K_{2i\nu}(\eta)= \frac{1}{4i\nu} \left [ K_{2i\nu+1}(\eta)-K_{2i\nu-1}(\eta)\right ] ,
\end{equation}
we evaluate the matrix element as
\begin{equation}
\langle \nu|e^{\xi/2}|\nu_1\rangle = \frac{1}{4}\left [ 
\frac{\langle \nu-i/2|\nu_1\rangle}{\sqrt{\nu(\nu-\frac{i}{2})}} + \frac{\langle \nu+i/2|\nu_1\rangle}{\sqrt{\nu(\nu+\frac{i}{2})}}
\right ]  . \label{MatElem}
\end{equation}
Combining Eqs~\eqref{Qdef}-\eqref{MatElem},  we obtain
\begin{align}
\mathcal{K}_{\alpha\uparrow\uparrow} &=   e^{-2 i\tilde{\epsilon}_{\alpha\uparrow} t_+} e^{\beta b/2} \frac{\sqrt{J^3}}{2\sqrt{\pi\beta}}
\int_{-\infty}^\infty  dh \sinh(h)  \notag \\
&\times 
 \sum_{p=\pm} e^{i pJ t_p/4} e^{i p \tilde{\epsilon}_{\alpha\uparrow} t_p}
 \mathcal{W}\bigl (2 h +i p J t_p, p\beta b/2, \beta J \bigr )
\notag \\
& \times 
\prod_{\gamma\neq\alpha} \prod_{\sigma=\pm} \left (1+e^{-\beta \tilde{\epsilon}_\gamma-\sigma h}\right ) . \label{K8}
\end{align}
Here we use the fact that $t_+-t_-=-i\beta$.  The function $\mathcal{W}$ is defined as
\begin{align}
\mathcal{W}(x,y,z) &= \frac{1}{4\sinh y} \Bigl [ \sum_{\sigma=\pm} \frac{\sigma \sqrt{\pi z} }{\sinh y} \erfi\left ( \frac{x-2\sigma y}{2\sqrt{z}}\right )
\notag \\
&+ 4 e^{-y} \, \exp\left (\frac{x-2 y}{2\sqrt{z}}\right )  \Bigr ]  ,
\end{align}
where  $\erfi(z) = (2/\sqrt\pi)\int_0^z dt\, \exp(t^2)$ is the error function of an imaginary argument. The expression for $\mathcal{K}_{\alpha\downarrow\downarrow}$ can be obtained from Eq.~\eqref{K8} by transforming $b\to -b$.

Finally, from Eqs~\eqref{CSSep1} and \eqref{CB} for $\tau>0$ we obtain 
\begin{align}
 G_{\alpha\uparrow\uparrow}(\tau) &= - \frac{1}{Z} 
\frac{\sqrt{\beta}}{\sqrt{\pi J}} e^{-J\tau/4} e^{\beta b/2} 
\sum_{n\in \mathbb{Z}} e^{-\beta E_c(n-N_0)^2}
\notag \\
&\hspace{-0.5cm} \times 
e^{-E_c(2n-2N_0+1)\tau} \int_{\pi/\beta}^{\pi/\beta} \frac{d\phi_0}{2\pi} \,
e^{i\beta \phi_0 n}  e^{- (\epsilon_{\alpha\uparrow}-\mu) \tau}\notag \\
&\hspace{-0.5cm} \times \int_{-\infty}^\infty  dh \sinh(h) \prod_{\sigma=\pm} \frac{e^{-\beta\Omega_0(\mu-i\phi_0+h \sigma/\beta)}}
{1+e^{-\beta(\epsilon_\alpha-\mu+i\phi_0)+h\sigma}}
\notag \\
&\hspace{-0.5cm}\times 
 \Biggl \{ 
e^{J (2\tau-\beta)/4} \mathcal{W}\bigl (2h+ J \tau, \beta b/2, \beta J\bigr )
\notag \\
&\hspace{-0.5cm}+  e^{- \beta(\epsilon_{\alpha\uparrow}-\mu+i\phi_0)}
 \mathcal{W}\bigl (2h+ J(\beta-\tau), -\beta b/2, \beta J\bigr )
\Biggr \} . \label{GUpUpF}
\end{align}

The integration over $\phi_0$ in Eq.~
\eqref{GUpUpF} can be performed with the help of the identity 
\begin{equation}
\prod_{\gamma\neq\alpha}  \left (1+e^{-\beta (\epsilon_\gamma-\mu)} \right ) = \sum_{N=0}^\infty Z_N(\epsilon_\alpha) e^{\beta \mu N} ,
\end{equation}
where $Z_N(\epsilon_\alpha)$ is the canonical partition function of a system of $N$
noninteracting spinless electrons under the constraint that level
$\alpha$ is not occupied. It is given by the Darwin-Fowler integral
\begin{equation}
Z_N(\epsilon_\alpha) \equiv\int_0^{2\pi} \frac{d\theta}{2\pi} e^{-i\theta N}
\prod_{\gamma\neq\alpha} \left( 1+e^{i\theta-\beta\epsilon_\gamma}\right) 
\end{equation}
and obeys the following identity: 
\begin{equation}
Z_N=Z_N(\epsilon_\alpha)+e^{-\beta\epsilon_\alpha} Z_{N-1}(\epsilon_\alpha). 
\end{equation}
Performing integration over $\phi_0$ and $h$ in Eq.~\eqref{GUpUpF}, we obtain (for $\tau>0$)
\begin{align}
G_{\alpha\uparrow\uparrow}(\tau) & = -\frac{1}{2Z} \sum_{n_{\uparrow,\downarrow}\in \mathbb{Z}}
e^{-\beta E_c(n-N_0)^2+\beta \mu n+ \beta Jm(m+1)]} \notag \\
&\hspace{-0.9cm} \times e^{-[\epsilon_{\alpha\uparrow} -\mu +E_c(2n-2N_0+1)+J(m+1/4)]\tau} \Biggl\{ e^{\beta b/2} \notag \\
&\hspace{-0.9cm} \times 
 \Upsilon(\beta b, 2m+1) \Bigl [ Z_{\nup}(\epsilon_\alpha)Z_{\ndn}-Z_{\nup+1}Z_{\ndn-1}(\epsilon_\alpha)\Bigr ]
\notag \\
&\hspace{-0.9cm} - \Upsilon(-\beta b,-2m) \Bigl [
Z_{\nup}Z_{\ndn}(\epsilon_\alpha)-Z_{\nup}(\epsilon_\alpha)Z_{\ndn}\Bigr ]  
\Biggr \} , \label{GK1}
\end{align}
where
\begin{equation}
\Upsilon(z,x)=\frac{e^{(x-1)z/2}}{\sinh(z/2)}-\frac{\sinh(x z/2)}{x \sinh^2(z/2)}  .
\end{equation}
We notice that $Y(z\to 0,x) = x-1$.
The expression for $G_{\alpha\downarrow\downarrow}(\tau)$ can be found from Eq.~\eqref{GK1} by reversing the sign of the magnetic field $b$. 

Employing  the general expression for the TDOS~\cite{MatveevAndreev}
\begin{equation}
\nu_\sigma(\ve) = -\frac{1}{\pi} \cosh \frac{\beta \ve}{2} \int\limits_{-\infty}^\infty dt\, e^{i\ve t}  
\sum\limits_{\alpha} G_{\alpha\sigma\sigma}\left ( it+\frac{\beta}{2}\right ) , 
\label{EqTDOSdef}
\end{equation}
finally, we find the following exact expression for the tunneling density of states for the UH~\eqref{EqUnivHam}:
\begin{align}
\nu_\sigma(\varepsilon) &= \frac{1+e^{-\beta \varepsilon}}{2Z} \sum_{n_{\uparrow},n_{\downarrow}} \frac{\sinh\frac{\beta b (2m+1)}{2}}{\sinh\frac{\beta b}{2}} 
Z_{\nup} Z_{\ndn} e^{\beta Jm(m+1)}
\notag \\
&\hspace{-0.7cm} \times e^{-\beta E_c(n-N_0)^2+\beta \mu n} \sum_\alpha \Biggl\{ 
\Biggl [ \frac{Z_{\nup}(\epsilon_\alpha)}{Z_{\nup}} +\frac{Z_{\nup}(\epsilon_\alpha)}{(2m+1)Z_{\nup}}\Biggr ] \notag \\
&\hspace{-0.63cm}\times \Bigl [ 1- B_{-m-1}\left (\sigma(m+1)\beta b\right )\Bigr ]\notag\\
 &\hspace{-0.63cm}\times
\delta\Bigl (\varepsilon-\epsilon_{\alpha\sigma} +\mu -E_c(2n-2N_0+1)+J(m+3/4)\Bigr ) 
\notag \\
&\hspace{-0.7cm} +
 \Biggl [ \frac{Z_{\ndn}(\epsilon_\alpha)}{Z_{\ndn}} -\frac{Z_{\nup}(\epsilon_\alpha)}{(2m+1)Z_{\nup}}\Biggr ]\Bigl [ 1+ B_m\left (\sigma m \beta b\right )\Bigr ]\notag\\
 &\hspace{-0.63cm}\times 
\delta\Bigl (\varepsilon-\epsilon_{\alpha\sigma} +\mu -E_c(2n-2N_0+1)-J(m+1/4)\Bigr )
\Biggr \}
  . 
\label{GK2} 
\end{align}
Here
\begin{equation}
B_m(x) = \frac{2m+1}{2m}\coth\left (\frac{2m+1}{2m}x\right ) - \frac{1}{2m} \coth \frac{x}{2m}
\end{equation}
denotes the Brillouin function. Equation~\eqref{GK2} constitutes the main result of the present paper. It allows to compute the TDOS for a given realization of single-particle levels. As expected, according to Eq.~\eqref{GK2}, the TDOS represents a sum of delta-functions corresponding to all possible processes of tunneling of an electron with energy $\varepsilon$ and spin $\sigma$ into (or from)  a single-particle level with energy $\epsilon_{\alpha\sigma}$. The factors $Z_n(\epsilon_\alpha)/Z_n$ describe the probability that the single-particle level $\alpha$ is empty.

By using the identity $\sum_\alpha [Z_n-Z_n(\epsilon_\alpha)]/Z_n = n$, one can check that the result~\eqref{GK2} satisfies the sum rule:  
\begin{equation}
\int_{-\infty}^\infty d\ve\,\frac{\nu_\sigma(\ve)}{1+\exp(\ve/T)} = -\sigma T 
\frac{\partial \ln Z}{\partial b} 
+\frac{T}{2} \frac{\partial \ln Z}{\partial\mu} . \label{SumRule}
\end{equation}

In the case $b=J=0$ and for spinless electrons the result~\eqref{GK2} coincides with the expression for the TDOS found in Ref.~[\onlinecite{SeldmayrLY}].


\section{Static spin susceptibility: the effect of level fluctuations and Zeeman splitting \label{Sec.SpinSusc}}

In this section we consider the thermodynamics of the quantum dot at relatively low temperatures, $\delta\ll T\ll E_c, E_{\rm Th}$. The quantity of main interest is the static spin susceptibility averaged over realizations of single-particle levels in the presence of Zeeman splitting. As is well established, its divergence indicates the Stoner instability. In general, the static spin susceptibility is defined as 
\begin{equation}
\chi(T,b) = T\frac{\partial^2 \ln Z}{\partial b^2} . \label{SpinSuscGen}
\end{equation}
In order to compute it one needs to perform integration over $\phi_0$ and $h$ in Eq.~\eqref{ZphiFinal_BFull}. At temperatures $T\gg \delta$, the integration over $\phi_0$ can be performed  in the saddle-point approximation.~\cite{KamenevGefen,EfetovTscherisch} Then, the partition function becomes 
\begin{equation}
Z = Z_C Z_S, \label{ZCZS}
\end{equation}
with
\begin{eqnarray}
Z_C &=& \sqrt{\frac{\beta\Delta}{4\pi}} \sum_{n\in\mathbb{Z}} e^{-\beta E_c(n-N_0)^2 +\beta \mu_0 n -2 \beta \Omega_0(\tilde\mu)} , \label{ZC1} \\
Z_S&=& \frac{1}{\sqrt{\pi \beta J}} e^{-\beta(J^2+b^2)/4J}
\int_{-\infty}^\infty dh\, \sinh( h) e^{-h^2/\beta J} \notag \\ 
&\times & \frac{\sinh(b h/J)}{\sinh(\beta b/2)}   \prod_\sigma e^{\beta [\Omega_0(\tilde\mu)-\Omega_0(\tilde\mu+h\sigma/\beta)]} . \label{ZS1}
\end{eqnarray}
Here $\tilde\mu=\mu+\mu_0$ with $\mu_0$ being the solution of the saddle-point equation: 
\begin{equation}
N_0 = -2 \frac{\partial \Omega_0(\mu+\mu_0)}{\partial \mu} .
\end{equation} 

This means that at $T\gg \delta$ the charging and the spin part of the problem are decoupled from each other. The partition function  in the absence of the exchange interaction and magnetic field is given by $Z_C$. The factor $Z_S$ in Eq.~\eqref{ZCZS} describes the effect of the exchange interaction and magnetic field: $Z_S=1$ at $J=b=0$. Next,
\begin{equation}
\Delta = - \left [ \frac{\partial^2\Omega_0(\tilde{\mu})}{\partial \tilde{\mu}^2}\right ]^{-1}
\end{equation}
stands for the inverse thermodynamic density of states at the Fermi level for a given realization of the single-particle spectrum. Since $Z_C$ is independent of the magnetic field and exchange interaction it does not affect the spin susceptibility. Therefore, in what follows we will discuss $Z_S$ only. In the absence of exchange interaction $\ln Z_S$ equals to $\beta \sum_\sigma [\Omega_0(\tilde\mu)-\Omega_0(\tilde\mu+b\sigma/2)]$ as it should.

The function $\beta\sum_\sigma [\Omega_0(\tilde{\mu})-\Omega_0(\tilde{\mu}+ h\sigma/\beta)]$ that appears in
Eq.~\eqref{ZS1} is a random function of the variable $h$ due to fluctuations in the single-particle density of states $\nu_0(E)=\sum_\alpha \delta(E+\tilde\mu-\epsilon_\alpha)$. Provided 
$h^2\ll \exp(\beta\tilde{\mu})$, we find
\begin{equation}
\beta\sum_\sigma \Bigl [\Omega_0(\tilde{\mu})-\Omega_0(\tilde{\mu}+ h\sigma/\beta)\Bigr ] = \frac{h^2}{\beta \delta}-V(h) ,\label{ZZeq2}
\end{equation} 
where 
 \begin{equation}
V(h) = -\int\limits_{-\infty}^\infty dE\, \delta\nu_0(E)\,  \ln \left [ 1+\frac{\sinh^2(h/2)}{\cosh^2(E/2T)}\right ] \label{Vh_Def}
\end{equation} 
is a random function. Here $\delta \nu_0(E)$ stands for the deviation of the single-particle density of states $\nu_0(E)$ from its average value:
\begin{equation}
1/\delta\equiv \overline{1/\Delta} .  \label{DeltaDefinition}
\end{equation}

With the help of Eq.~\eqref{ZZeq2} we rewrite Eq.~\eqref{ZS1} as
\begin{gather}
Z_S = \frac{1}{\sqrt{\pi J \beta}} \frac{\Xi(b/J,\beta J_\star)}{\sinh(\beta b/2)} \exp \left (
-\beta \frac{J^2+b^2}{4J} \right )
 ,\notag  \\
\Xi(x,y)= \int_{-\infty}^\infty dh\, \sinh (x h) e^{h-h^2/y -V(h)} . \label{ZS1_3}
\end{gather}
Here 
\begin{equation}
J_\star\equiv \frac{J \delta}{\delta-J}
\label{JstarDef}
\end{equation}
denotes the renormalized exchange energy. We emphasize that near the Stoner instability, $\delta-J\ll \delta$, the value of spin  
in the ground state is of the order of $\delta/[2(\delta-J)]$,~\cite{KAA} and the renormalized exchange energy is much larger than the bare exchange, $J_\star \gg J$. In what follows, we consider this most interesting regime. 

Let us neglect $V(h)$ in Eq.~\eqref{ZS1_3} for a moment. Then, the typical value of $h$ in the integral of Eq.~\eqref{ZS1_3} is of the order of $\max\{\sqrt{y},y,yx\}$. Therefore, Eq.~\eqref{ZZeq2} is valid if the following condition holds: $\beta J_\star\max\{1,1/\sqrt{\beta J_\star}, b/J\}\ll\exp(\beta\tilde{\mu})$. It is satisfied for large enough values of the chemical potential.

Although the single-particle density of states $\nu_0(E)$ has non-Gaussian statistics, for $\max\{|h|,T\}\gg \delta$
the function $V(h)$ is a Gaussian random variable.~\cite{Mehta} Its statistics is fully determined by the pair correlation function
\begin{equation}
C(h_1,h_2) =\overline{V(h_1)V(h_2)} . \label{DefCorr_0}
\end{equation} 
The following exact relation holds (see Appendix~\ref{App_CorrFunc}):
\begin{equation}
C(h_1,h_2)  = L(h_1+h_2)+L(h_1-h_2) - 2 L(h_1) - 2L(h_2) . \label{DefCorr}
\end{equation}
At $T\gg\delta$ the behavior of the correlation function $L$ depends strongly on the value of $|h|$:
\begin{equation}
L(h) = \frac{h^2}{\pi^2\bm{\beta}}\begin{cases}
c_1 h^2/24,\quad |h|\ll 1 ,\\ 
\ln(|h|/2)+c_2-3/2, \quad |h|\gg 1 .
\end{cases} \label{Sassymp}
\end{equation}
Here we introduce the parameter $\bm{\beta}$ such that $\bm{\beta}=2$ 
for the unitary Wigner-Dyson ensemble (class A) and $\bm{\beta}=1$ for the orthogonal Wigner-Dyson ensemble (class AI). 
The numerical constants are 
\begin{align}
c_1 &=  \int_0^\infty \frac{d\omega}{\omega^2} \Biggl \{
\frac{1}{3} - \frac{\omega\coth\omega -1}{\sinh^2\omega} 
 \Biggr \} \approx 0.37 , \\
c_2 &= -\int_0^1\frac{d\omega}{\omega^2}[1-\omega\coth\omega]+\int_1^\infty \frac{d\omega\,\ln\omega}{\sinh^2\omega} \approx 0.43 .
\label{Eq_Def_c2}
\end{align}

\begin{figure}[t]
\centerline{\includegraphics[width=75mm]{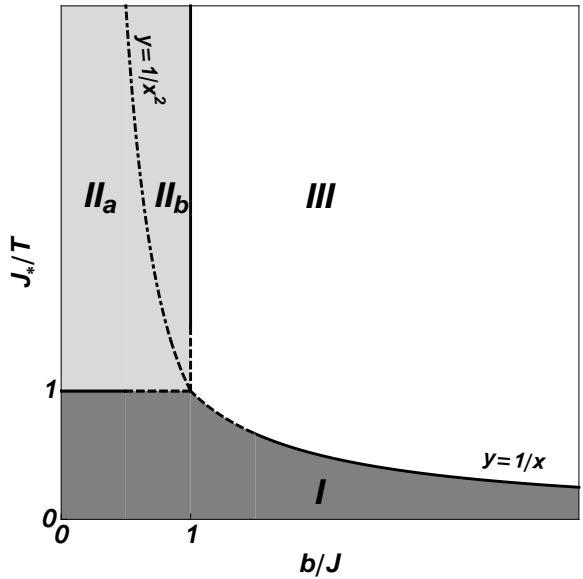}}
\caption{Different regions for behavior of the spin susceptibility in the plane of dimensional parameters $b/J$ and $J_\star/T$. Note that in our analysis $T\gg \delta$.}
\label{FigureRegions}
\end{figure}

In spite of the fact that $V(h)$ is a Gaussian random variable exact evaluation of $\overline{\ln \Xi(x,y)}$ for arbitrary values of $x$ and $y$ is a complicated problem (see e.g., Ref.~[\onlinecite{Fyodorov}]). In what follows, we compute $\overline{\ln \Xi(x,y)}$ to the first order in the correlation function $C$. We expand   
expression~\eqref{ZS1_3} for $\Xi(x,y)$ to the second order in $V$ and perform averging of $\ln \Xi(x,y)$ with the help of Eq.~\eqref{DefCorr}. Then, we find
\begin{align}
\overline{\ln \Xi(x,y)} &= \ln \Xi_0(x,y)  + \frac{1}{4}\sum_\sigma \int_{-\infty}^\infty \frac{du}{\sqrt\pi} e^{-u^2} \notag \\
& \times \Biggl \{
\Bigl [ \sigma  \coth \frac{xy}{2} +1\Bigr] \Bigl [ L\bigl (y(1+x\sigma)+2u\sqrt{y}\bigr)\notag \\
&- L\bigl (y(1+x\sigma)+u\sqrt{2y}\bigr)\Bigr ] - L\bigl (u\sqrt{2y}\bigr ) \notag \\
&- \frac{1}{2\sinh^2(xy/2)}\Bigl [L\bigl ( y(1+x\sigma)+u\sqrt{2y}\bigr)\notag \\
&-L\bigl ( y+u\sqrt{2y}\bigr) -L\bigl ( y x\sigma+u\sqrt{2y}\bigr)\notag \\
&+L\bigl ( u\sqrt{2y}\bigr) \Bigr] 
\Biggr \} , \label{AverLnXi1}
\end{align}
where 
\begin{equation}
\Xi_0(x,y) = \sqrt{\pi y}\, e^{y(1+x^2)/4} \sinh\frac{xy}{2}  .
\end{equation}
Exact integration over $u$ in Eq.~\eqref{AverLnXi1} is complicated since only asymptotic expressions~\eqref{Sassymp} for $L$ are known. To make further analytical progress it is useful to consider separately regions in which arguments of the functions $L$ involved in Eq.~\eqref{AverLnXi1} are either large or small. This way one finds several regions shown in 
Fig.~\ref{FigureRegions}; in each of those the behavior of $\overline{\ln \Xi(x,y)}$ is different.

%
\subsection{Region I: $J_\star \max\{1,b/J\} \ll T$ \label{Sec_RegI}}

In the region I, $\max\{y, xy\}\ll1$, the arguments in all functions $L$ in the right hand side of Eq.~\eqref{AverLnXi1} are much smaller than unity. 
Therefore, in order to evaluate integrals over $u$ we can use the asymptote of $L(h)$ for $|h|\ll1$ (see Eq.~\eqref{Sassymp}). Then, we obtain
\begin{align}
\overline{\ln \Xi(x,y)} & = \ln \Xi_0(x,y) + \frac{c_1}{96\bm{\beta}\pi^2} \Bigl [ 30 y^2 +12 y^3 \notag \\
& + 12 x^2 y^3  (1+y/2) - x^4 y^6/6 \Bigl ] .
\end{align}
Hence, using Eqs~\eqref{SpinSuscGen} and \eqref{ZS1_3}, we find the average spin susceptibility in the region I ($J_\star \max\{1,b/J\} \ll T$):
\begin{align}
\overline{\chi(T,b)} & =  \frac{J_\star}{2J\delta} \Biggl \{1 + \frac{J_\star}{6T} +\frac{c}{\bm{\beta}} \frac{J_\star^2}{T^2}\Bigl [ 1 +\frac{J_\star}{2T}\Bigr ]
\notag \\
& - \frac{J_\star^3 b^2}{120 T^3 J^2} \Bigl [ 1 + \frac{10 c}{\bm{\beta}} \frac{J_\star^2}{T^2}\Bigr ] \Biggr \} . 
\label{SpinSuscRegI_Fin}
\end{align}
Here the numerical constant $c = c_1/2\pi^2 \approx 0.02$. 

It is instructive to derive Eq.~\eqref{SpinSuscRegI_Fin} in a more transparent way. 
In region I the typical value of $h$ in the integral in Eq.~\eqref{ZS1_3} is much smaller than unity. Therefore, we can expand the random function $V(h)$ as a series in $h$. We thus find:
\begin{equation}
V(h) \approx \left ( \frac{1}{\Delta}-\frac{1}{\delta}\right ) T h^2, \qquad |h|\ll 1 . \label{Vh_small}
\end{equation}
Next, performing integration over $h$ in Eq.~\eqref{ZS1_3}, we obtain
\begin{equation}
Z_S = \frac{\sqrt{\mathcal{J}}}{\sqrt{J}} \frac{\sinh [\mathcal{J} b/(2JT)]}{\sinh(b/2T)}e^{(\mathcal{J}-J)(J^2+b^2)/4J^2T}  ,\label{ZS2}
\end{equation}
where $1/\mathcal{J} = 1/J -1/\Delta$ stands for the renormalized exchange interaction for a given realization of single-particle spectrum. All information about level statistics is contained in fluctuations of $\Delta$. At $T\gg \delta$ the fluctuations of $\Delta$ are small and Gaussian with (see Appendix~\ref{App_CorrFunc})
\begin{equation}
\overline{(\Delta-\delta)^2} = \frac{c}{\bm{\beta}} \frac{\delta^4}{T^2} . \label{FlucRes}
\end{equation}

To avoid the Stoner instability the renormalized exchange energy $\mathcal{J}$ should be positive for a given realization of single-particle spectrum. Since we are interested in the regime $J_\star \gg \delta \gtrsim J$, in order to fulfill the condition $\mathcal{J}>0$ the level fluctuations should be such that  
\begin{equation}
\overline{\left ( \frac{1}{\Delta}-\frac{1}{\delta}\right )^2} \ll \overline{\left (\frac{1}{\mathcal{J}}\right )^2} . \label{condR1}
\end{equation}
Using Eq.~\eqref{FlucRes}, we find that Eq.~\eqref{condR1} is equivalent to the condition $T\gg J_\star$. Provided  the latter is satisfied (i.e in the region I) the level fluctuations are small enough and cannot drive the system to be Stoner unstable. 

Expanding Eq.~\eqref{ZS2} to the forth order in $b$ and performing averaging over level fluctuations by means of the expression $\overline{\mathcal{J}^n} = J^n_\star [1+n(n+1)c J^2_\star/(2 \bm{\beta} T^2)]$ we obtain  Eq.~\eqref{SpinSuscRegI_Fin}.
The expression~\eqref{SpinSuscRegI_Fin} underlining  the divergence at the Stoner
instability point represents exchange-enhanced Pauli spin-susceptibility ($J_\star/2J\delta$) with small corrections depending on temperature and magnetic field. The corrections due to level fluctuations are small. 

It is worthwhile to mention that the spin susceptibility at $b=0$ has been studied by Kurland et al.~\cite{KAA} In our notations, their result at $T\gg J_\star$ can be written as
(see Eqs (4.8), (4.13b), (4.15) of Ref.~[\onlinecite{KAA}]):
\begin{gather} 
\overline{\chi_{KAA}(T,0)}   = \frac{J_\star}{2 J\delta}\left [  \frac{2}{3} + \frac{\sqrt{\pi}}{6} \frac{\sqrt{J_\star}}{\sqrt{T}} + c_{KAA} \frac{J_\star}{T} \right] , \label{KAA_exp} \\
c_{KAA} = \frac{4-\pi}{6} + \frac{1}{6\bm{\beta}} \Bigl ( \frac{8 \ln 2}{\pi^2} - 0.3712\Bigr )  .
 \notag 
\end{gather}
The result~\eqref{KAA_exp} of Ref.~[\onlinecite{KAA}]
 disagreees with our result~\eqref{SpinSuscRegI_Fin}.

\subsection{Region II: $\delta \ll T\ll J_\star$ and $b \ll J$ \label{Sec_RegII}}

In the region II, $y\gg 1$ and $x\ll 1$, the arguments of all functions $L$ in Eq.~\eqref{AverLnXi1} are typically much larger than unity. However, the behavior of some contributions to the integral over $u$ in Eq.~\eqref{AverLnXi1} are different for $x^2y\gg 1$ and $x^2y\ll 1$. Therefore, it is convenient to split region II into two regimes.

\subsubsection{Region II$_a$: $\delta \ll T\ll J_\star$ and $b^2/J^2 \ll T/J_\star$}

Let us first consider the regime of weak magnetic fields: $x^2\ll 1/y\ll 1$ (region II$_a$). In this case, we  perform integration over $u$ in Eq.~\eqref{AverLnXi1} either by expansion of $L$ to second order in $u$ or with the help of the following asymptotic result at $z\gg 1$:
\begin{equation}
\int_{-\infty}^\infty \frac{du}{\sqrt{\pi}} e^{-u^2} L(z u) \approx \frac{z^2}{2\bm{\beta}\pi^2}\Bigl [\ln \frac{z}{4}+c_2-\frac{1+\gamma}{2} \Bigr ]  .
\label{Lasymp}
\end{equation}
Here and below, $\gamma \approx 0.577\dots $ stands for the Euler's constant, and constant $c_2$ is defined in Eq.~\eqref{Eq_Def_c2}.
We obtain
\begin{align}
\overline{\ln \Xi(x,y)} &= \ln \Xi_0(x,y)  +\frac{1}{2\bm{\beta}\pi^2} \Bigl [ xy \coth\frac{xy}{2} \notag \\
&- \frac{x^2y^2}{4\sinh^2(xy/2)} (\ln 2y+\gamma)\Bigr ] . \label{lnXi2}
\end{align}
Using Eqs~\eqref{ZS1_3} and~\eqref{lnXi2}, we find the average spin susceptibility (defined in Eq.~\eqref{SpinSuscGen}) in the region II$_a$ ($b^2/J^2\ll T/J_\star\ll 1$):
\begin{align}
\overline{\chi(T,b)} & = \frac{T}{b^2} \Bigl [ 1-\frac{J_\star^2 b^2}{4 J^2T^2 \sinh^2(J_\star b/2 J T)}\Bigr ] \notag \\
&+ \frac{T}{2\bm{\beta}\pi^2} \frac{\partial^2}{\partial b^2} \Biggl [ \frac{J_\star b}{JT} \coth \frac{J_\star b}{2T J} \notag \\
& -  \frac{J_\star^2 b^2\Bigl ( \ln (2 J_\star/T)+\gamma\Bigr )}{4 J^2T^2\sinh^2(J_\star b/2 J T)}  \Biggr ] . 
\label{SpinSuscRegIIa_Fin}
\end{align}

In the limit of very weak magnetic fields $b \ll J T/J_\star$ we can neglect tiny dependence of the average spin susceptibility on magnetic field. Then, from Eq.~\eqref{SpinSuscRegIIa_Fin} we obtain the following result for the zero-field average spin susceptibility:
\begin{equation}
\overline{\chi(T,0)} = \frac{(J_\star/J)^2}{12 T}\left \{1+ \frac{1}{\bm{\beta}\pi^2}\left [ \ln \frac{2J_*}{T} + \gamma+2\right ]\right \} .\label{chi_fin_f}
\end{equation}
The result~\eqref{chi_fin_f} is valid provided the expansion of $\overline{\ln \Xi(x,y)}$ in powers of the correlation function~\eqref{DefCorr_0} is justified.  As one can demonstrate (see Appendix~\ref{Appendix_Replica}), the latter is controlled by the small parameter $J_*/(\bm{\beta}\pi^2 T) \ll 1$. Therefore, strictly speaking, the result~\eqref{chi_fin_f} holds at $1\ll J_\star/T \ll \bm{\beta}\pi^2$. However, the more detailed analysis of $\overline{\chi(T,0)}$ presented in Appendix~\ref{Appendix_Replica}, together with the qualitative arguments of Sec.~\ref{Sec_Chi_Qual} below, allow us to speculate that the result~\eqref{chi_fin_f} has a much broader range of applicability. 

We expect that the zero-field average spin susceptibility at $J_\star\gg T \gg \delta$ can be written as
\begin{equation}
\overline{\chi(T,0)} = \frac{(J_\star/J)^2}{12 T}\left \{1+ \frac{1}{\bm{\beta}\pi^2}\left [ \ln \frac{J_*}{T} +  f\Bigl (\frac{J_*}{\bm{\beta}\pi^2T}\Bigr)\right ] \right \} .\label{chi_fin_f1}
\end{equation}
where $f(x)$ is constant of the order unity in both limiting cases of
small ($x\ll 1$) and large ($x\gg 1$) values of its argument. In
particular, Eq.~\eqref{chi_fin_f} implies that $f(x) = \gamma+2+\ln 2
+ \cdots$ at $x\ll 1$.

Therefore, near the Stoner instability, $\delta-J\ll\delta$, there is ehnancement of the average spin susceptibility (Eq.~\eqref{chi_fin_f})
due to fluctuations of single-particle levels in a wide temperature range $\delta\ll T \ll J_\star$.

In the regime of larger magnetic fields $J T/J_\star\ll b \ll \sqrt{J T/J_\star}$ from Eq.~\eqref{SpinSuscRegIIa_Fin} we find the following result for the average spin susceptibility:
 \begin{align}
\overline{\chi(T,b)} = \frac{T}{b^2} \Biggl [ 1 - \frac{J_\star^2 b^2}{J^2 T^2} \Bigl (1 + \frac{1}{2\bm{\beta}\pi^2} \frac{J_\star^2 b^2}{J^2 T^2}  \Bigr )
e^{-J_\star b/J T}\Biggr ] .
\end{align}
The spin susceptibility is suppressed as compared to the zero-field result~\eqref{chi_fin_f}. It is given mainly by $T/b^2$ term whereas the level fluctuations contribute to the terms which are exponentially small, $\propto\exp(-J_\star b/J T)$.

\subsubsection{Region II$_b$: $\delta \ll T\ll J_\star$ and $J_\star/T\ll b^2/J^2\ll 1$ }

Let us now consider the regime of intermediate magnetic fields: $1/y\ll x^2\ll 1$ (region II$_b$). Performing integration over $u$ in Eq.~\eqref{AverLnXi1}, as in the previous section,  either by expansion of $L$ to the second order in $u$ or using the asymptotic result~\eqref{Lasymp} we obtain
\begin{equation}
\overline{\ln \Xi(x,y)} = \ln \Xi_0(x,y)  +\frac{x^2 y^2}{\bm{\beta}\pi^2}
\left (\ln x-\frac{3}{2}\right )e^{-xy} . \label{lnXi3}
\end{equation}
With the help of Eqs~\eqref{ZS1_3} and~\eqref{lnXi3}, we find that the average spin susceptibility in the region II$_b$ ($T/J_\star\ll b^2/J^2\ll 1$) is given as
\begin{align}
\overline{\chi(T,b)} & = \frac{J_\star}{2J\delta} \Biggl \{ 1-\frac{2J_\star}{T} e^{- J_\star b/J T}
\notag \\
& + \frac{2 J_\star}{\bm{\beta}\pi^2 T} \Bigl (\ln\frac{b}{J} - \frac{3}{2}\Bigr ) e^{- J_\star b/J T}
\Biggr \}. \label{SpinSuscRegIIb_Fin}
\end{align}
We note that a small magnetic field $b \sim (J/J_\star) T\ll J, T$ destroys the phenomenon of mesoscopic Stoner instability, i.e. the existence of the ground state of a quantum dot with a finite non-zero spin. It renders the spin-susceptibility of the Fermi-liquid type. Since 
$J_\star/J$ is a factor responsible for the enhancement of a $g$-factor in the Fermi-liquid, the energy scale  
$J_\star b/J$  determines the effective Zeeman splitting. As usual, comparison of the latter with temperature allows us to distinguish between weak and strong magnetic fields.

\subsection{Region III: $\delta\ll T\ll b J_\star/J$ and $J\ll b$\label{RegIIISpinSusc}}

Similarly to the region II, in the region III ($y\gg 1/x$ and $x\gg 1$) the arguments of all functions $L$ in Eq.~\eqref{AverLnXi1} are typically much larger than unity. However, in the region III for evaluation of the integral in Eq.~\eqref{AverLnXi1} one can expand all functions $L$ to the second order in $u$ 
except $L(u\sqrt{2y})$. The latter can be integrated by means of Eq.~\eqref{Lasymp}. Then, we obtain
\begin{equation}
\overline{\ln \Xi(x,y)} = \ln \Xi_0(x,y)  +\frac{y}{2 \bm{\beta}\pi^2} \Bigl ( \ln \frac{xy}{2} +c_2\Bigr ). \label{lnXi4}
\end{equation}
Hence, using Eqs~\eqref{ZS1_3}, we find the average spin susceptibility (defined in Eq.~\eqref{SpinSuscGen}) in the region III ($T/J_\star\ll b/J$ and $J\ll b$):
\begin{equation}
\overline{\chi(T,b)}  = \frac{J_\star}{2J\delta} \left (1- \frac{1}{\bm{\beta}\pi^2}\frac{J^2}{b^2} 
\right ). \label{chi_fin_f_r3}
\end{equation}
We recall that in the course of the derivation of Eq.~\eqref{chi_fin_f_r3} we have neglected terms exponentially small in $b$. 
The spin susceptibility~\eqref{chi_fin_f_r3} is exchange-enhanced Pauli susceptibility, with correction due to fluctuations of single-particle levels. The latter is small in the regime considered. It suggests that taking the lowest order expansion in $V(h)$ for evaluating $\overline{\ln \Xi(x,y)}$ is well-justified in the whole region III. We mention that our result~\eqref{chi_fin_f_r3} for the spin susceptibility is an extention of the zero-temperature result of Ref.~[\onlinecite{Schechter}] with account for the effect of level fluctuations.

It is instructive to summarize here the results of the above analysis of the average spin susceptibility. In region I (see Fig.~\ref{FigureRegions}), the mesoscopic Stoner instability is manifested through small temperature and magnetic field dependent corrections to the Fermi-liquid result. The effect of level statistics is weak. In regions II$_b$ and III, the mesoscopic Stoner instability is suppressed by magnetic field. The corrections to the Fermi-liquid result due to fluctuations of the single-particle levels are small. In region II$_a$, the average spin susceptibility behaves in accordance with the Curie law. The latter is a manifestation of the mesoscopic Stoner instability at non-zero temperature. Fluctuations of single-particle levels lead to logarithmic-in-temperature corrections to the Curie term in the average spin susceptibility. 

Here we do not study fluctuations of spin susceptibility due to statistical fluctuations of single-particle levels. However, we expect that in all regions except region II$_a$ they are small. In region II$_a$, we estimate $(\overline{\chi^2}-(\overline{\chi})^2)/(\overline{\chi})^2$ to be of the order of $(\ln J_\star/T)/(\bm{\beta}\pi^2)$. This result indicates that in region II$_a$ fluctuations of the spin susceptibility can be large and, therefore, it is challenging to study the whole distribution function for the spin susceptibility.

\section{Semi-qualitative analysis of the effect of the level fluctuations on the spin-susceptibility at $T=0$ 
\label{Sec_Chi_Qual}}

\subsection{Spin susceptibility at zero magnetic field}

The origin of the logarithm in the result~\eqref{chi_fin_f} has simple physical explanation. Although the result~\eqref{chi_fin_f}  was derived for $T\gg \delta$, let us consider the limiting case of vanishing temperature, $T=0$. Then, the transition between the ground states with the spin $S$ and $S+1$ occurs if 
\begin{equation}
E_{S+1}-E_S = J(2S+2) , \label{Crit1}
\end{equation}
where $E_S$ stands for the single-particle contribution to the ground state energy (`kinetic energy')
for a given realization of single-particle levels. It can be estimated as 
\begin{equation}
E_{S+1}-E_S = \delta (2S+1) +\Delta E_{2S} . \label{Crit2_0}
\end{equation}
Here $\Delta E_{2S}$ is a fluctuation of the energy strip in which there are $2S$ levels in average. We can estimate it as
\begin{equation}
\Delta E_{2S}= \delta \, \Delta n_{2S} , \label{DeltaE_Def}
\end{equation}
 where $\Delta n_{2S}$ is a fluctuation of the number of levels in the energy strip $\Delta E_{2S}$. Near the Stoner instability, we find
\begin{equation}  
S = \frac{\delta}{2(\delta-J)} \Bigl [ 1 - \Delta n_{2S}\Bigr ] . \label{Sfluc}
\end{equation}
Hence, the average spin susceptibility at low temperatures can be esimated as
\begin{equation}
\overline{\chi(T,0)} = \frac{\overline{S(S+1)}}{3T}  \sim \frac{J_\star^2}{12 T J^2} \Bigl [ 1+\overline{(\Delta n_{J_\star/J})^2}\Bigr ]
\end{equation}
From random matrix theory it is well-known that~\cite{Mehta} 
\begin{equation}
\overline{(\Delta n_{2S})^2} = \frac{2}{\bm{\beta} \pi^2} \Bigl [ \ln 2S +{\rm \, const}\Bigr ] . \label{Crit2}
\end{equation}
Hence we find the following estimate for the spin susceptibility  at $T\ll \delta$:
\begin{equation}
\overline{\chi(T,0)}  \sim \frac{J_*^2}{12 T J^2} \Bigl [ 1+\frac{2}{\bm{\beta} \pi^2} \Bigl (\ln \frac{J_*}{J}+{\rm const}\Bigr )\Bigr ] .
\label{chi_fin_f_T0}
\end{equation}
The estimate~\eqref{chi_fin_f_T0} derived from qualitative arguments for $T\ll \delta$ resembles the result \eqref{chi_fin_f} which is
valid at higher temperatures $T\gg\delta$. There is a discrepancy in factor of $2$ in front of the logarithm between estimate~\eqref{chi_fin_f_T0} and the rigorous result~\eqref{chi_fin_f}. The reason for this discrepancy is the following. The qualitative argumens presented above correspond to evaluating the integral over $h$ in Eq.~\eqref{ZS1_3} in the saddle-point approximation. In such a procedure one misses the contributions of the type $L(u\sqrt{2y})$ in Eq.~\eqref{AverLnXi1}. 

\begin{figure}[t]
\centerline{\includegraphics[width=75mm]{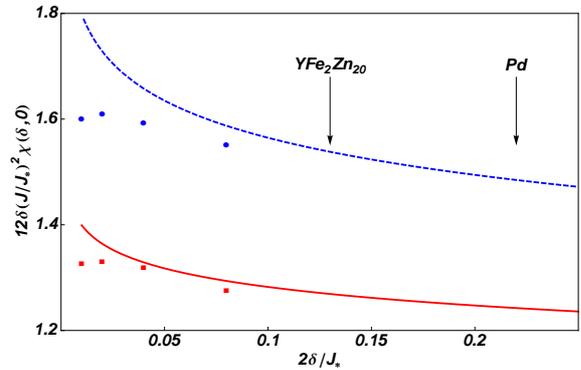}}
\caption{(Color online) Comparison between the average spin susceptibility at zero temperature and zero magnetic field estimated from
the theoretical result~\eqref{chi_fin_f} with $T=\delta$ (solid curve for $\bm{\beta}=2$ and dashed curve for $\bm{\beta}=1$) and results of numerical simulations of Ref.~\protect[\onlinecite{KAA}] (circles for $\bm{\beta}=2$ and squares for $\bm{\beta}=1$).}
\label{FigureBGKvsKAA}
\end{figure}

In Fig.~\ref{FigureBGKvsKAA} we present the comparison between the average spin susceptibility at zero temperature and zero magnetic field estimated from Eq.~\eqref{chi_fin_f} with $T\sim \delta$ and results of numerical simulations of Ref.~\protect[\onlinecite{KAA}].

\subsection{Spin susceptibility at a strong magnetic field, $b\gg J$}

The result~\eqref{chi_fin_f_r3} can be illustrated by simple physical argumentation. As above, we consider the limiting case of vanishing temperature, $T=0$. Then, the difference between the single-particle contributions to the ground state energy in Eq.~\eqref{Crit1} 
can be estimated as 
\begin{equation}
E_{S+1}-E_S = \delta (2S+1) - b +\Delta E_{2S} .
\end{equation}
Near the Stoner instability, we find
\begin{equation}  
S = \frac{1}{2(\delta-J)} \Bigl [ b - \delta\, \Delta n_{2S}\Bigr ] . \label{CritS3}
\end{equation}
Hence, the average spin susceptibility at low temperatures can be estimated as
\begin{equation}
\overline{\chi(T,b)} \sim \frac{\partial \overline{S}}{\partial b}  = 
\frac{J_\star}{2 J\delta} \left [ 1 + \frac{J_\star^2}{2 \delta^2} \frac{d^2 \overline{(\Delta n_z)^2}}{d z^2}\right ]\Biggr |_{z=J_\star b/J \delta}.
\end{equation}
Using Eq.~\eqref{Crit2}, one exactly recovers the result~\eqref{chi_fin_f_r3}. Since our result~\eqref{chi_fin_f_r3} can be obtained from zero-temperature considerations presented above, we expect Eq.~\eqref{chi_fin_f_r3} to be valid at $T=0$ provided $b\gg J, \delta$.

%
%

\section{Tunneling density of states in magnetic field in the absence of level fluctuations\label{Sec.TDOS.An}}

In this section we analyze the TDOS at sufficiently low temperatures $\delta \ll T\ll E_c, E_\textrm{Th}$. As 
it was demonstrated in the previous section the effect of exchange interaction is most pronounced in the vicinity of the Stoner instability, $\delta-J\ll \delta$. Therefore, we shall consider the regime $J_\star \gg J$ below. As was indicated in our above analysis of the spin susceptibility, here, too, there are the same three regions with different dependence of the TDOS on temperature and magnetic field (see Fig.~\ref{FigureRegions}). The fluctuations of single-particle levels are important in region II$_a$ only. Most of our discussion below excludes the effect of level fluctuations. We eventually present semi-qualitative arguments to account for the effect of the latter.


We start from rewritting Eq.~\eqref{GUpUpF} as 
\begin{align}
& G_{\uparrow\uparrow}(\tau) = \frac{1}{2 Z} 
\frac{\sqrt{\beta}}{\sqrt{\pi J}} e^{-J\tau/4} e^{(\beta-\tau) b/2} 
\sum_{n\in \mathbb{Z}} e^{-\beta E_c(n-N_0)^2}
\notag \\
& \times 
e^{-E_c(2n-2N_0+1)\tau +i \phi_0\tau} \int_{\pi/\beta}^{\pi/\beta} \frac{d\phi_0}{2\pi} 
e^{i\beta \phi_0 n}   \int_{-\infty}^\infty  dh
\notag \\
& \times  \left [ \prod_{\sigma=\pm} e^{-\beta\Omega_0(\mu-i\phi_0+h \sigma/\beta)} \right ]
G_0(\tau, \mu-i\phi_0+h/\beta) \notag \\
&\times
 \Biggl \{ 
e^{J (2\tau-\beta)/4} e^{(1-\tau/\beta)h} \sum_{p=\pm} p \mathcal{W}\bigl (2 p h+ J \tau, \beta b/2, \beta J\bigr )
\notag \\
&-  e^{- \beta b/2} e^{-h \tau/\beta} 
\sum_{p=\pm} p  \mathcal{W}\bigl (2 p h+ J(\beta-\tau), -\beta b/2, \beta J\bigr )
\Biggr \} . \label{GUpUpF1}
\end{align}
Here the Green's function of non-interacting electrons in imaginary time is given by 
\begin{gather}
G_0(\tau, \mu) = -\int dE \, \frac{ \nu_0(E)}{2\cosh(\beta E/2)} e^{E(\tau-\beta/2)} . \label{GreenNonInt}
\end{gather}
Provided the condition $\delta\ll T\ll \mu$ holds, the  Green's function~\eqref{GreenNonInt} can be simplifed to
\begin{gather}
G_0(\tau,\mu) = -(\pi T/\delta)/\sin(\pi T\tau) . 
\end{gather}
Integration over $\phi_0$ and $h$ in Eq.~\eqref{GUpUpF1} can be performed in the same way as it was done in the previous section for derivation of the 
spin susceptibility. We remind that integration over $h$ was done under assumption that 
$\beta J_\star\max\{1,1/\sqrt{\beta J_\star}, b/J\}\ll\exp(\beta\tilde{\mu})$. Then, using Eq.~\eqref{ZCZS}, we obtain
\begin{align}
G_{\uparrow\uparrow}(\tau) &= - \frac{\pi/(\beta\delta)}{\sin(\pi\tau/\beta)} \frac{\sum\limits_{n\in \mathbb{Z}} e^{-\beta E_c(n-N_0+\tau/\beta)^2}
 }{\sum\limits_{n\in \mathbb{Z}} e^{-\beta E_c(n-N_0)^2}}e^{E_c\tau(\tau/\beta-1)}
\notag \\
\times &
\Biggl \{
1+\frac{\sqrt{\pi\beta} J e^{(\beta-2\tau) b/4}
e^{-\beta J_\star b^2/4J^2}e^{-\beta J_\star/4}}{8\sqrt{J_\star} \cosh(\beta b/4)\sinh(J_\star \beta b/2J)}  
\notag \\
\times &
\sum_{\sigma=\pm} \sigma 
\Bigl [ 
\erfi\bigl (\sqrt{\beta J_\star}(\sigma b-J)/2J\bigr ) \notag \\
 & -  
\erfi\bigl (\sqrt{\beta J_\star}(\sigma b+J(1-2\tau/\beta))/2J\bigr ) 
\Bigr ] 
\Biggr \}
 . \label{GUpUpF2}
\end{align}
The Green's function $G_{\uparrow\uparrow}(\tau)$ determines the TDOS in accordance with Eq.~\eqref{EqTDOSdef}. Performing 
integration over $t$ 
with the help of the following identity:
\begin{gather}
\int_{-\infty}^\infty dt \frac{e^{i z t}}{\cosh(\pi t)} = \frac{1}{\cosh(z/2)} ,
\end{gather}
we find
\begin{align}
\frac{\nu_\sigma(\varepsilon)}{\nu_0} &= \sum\limits_{n\in \mathbb{Z}} e^{-\beta E_c(n-N_0)^2}
\sum_{p=\pm}  \Biggl \{ f_F\bigl ( p\ve - 2 p \Omega_n^{-p}\bigr )  \notag \\
&+\frac{\sqrt{\pi \beta} J e^{\sigma \beta b/4}
e^{-\beta J_\star b^2/4J^2}e^{-\beta J_\star/4}}{8\sqrt{J_\star} \cosh(\beta b/4)\sinh(J_\star \beta b/2J)}  
\sum_{s=\pm} s  \notag \\
&\times  \Bigl [ 
\erfi\Bigl (\frac{\sqrt{\beta J_\star}(s b-J)}{2J}\Bigr ) f_F\Bigl ( p\ve - 2 p \Omega_n^{-p}+\frac{\sigma b}{2}\Bigr ) \notag \\
&- 
\mathbb{F}\Bigl (\beta (p\ve - 2 p \Omega_n^{-p}+\sigma b/2),s b/2J,\sqrt{\beta J_\star}\Bigr )
\Bigr ] \Biggr \}\notag \\
&\times  \left [ \sum_{n\in \mathbb{Z}} e^{-\beta E_c(n-N_0)^2} \right ]^{-1}. \label{FinalNu}
\end{align}
Here $\nu_0\equiv 1/\delta$ denotes the average density of states of non-interacting electrons for single spin projection, $\Omega_n^p = E_c(n-N_0+p/2)$, $f_F(\ve)=
[1+\exp(\ve/T)]^{-1}$ stands for the Fermi function, 
and the function
\begin{gather}
\mathbb{F}(x,y,z)= \frac{1}{2}e^{-x/2}\int_{-\infty}^\infty dt\, \frac{e^{ix t}}{\cosh(\pi t)} \erfi\bigl( z (y-i t)\bigr )  .
\end{gather}
Equation~\eqref{FinalNu} describes dependence of the tunneling density of states on energy, temperature and magnetic field. It is valid under the following conditions: $\delta\ll T\ll \mu$ and $\beta J_\star\max\{1,1/\sqrt{\beta J_\star}, b/J\}\ll\exp(\beta\tilde{\mu})$, and near the Stoner instability,  
$J_\star\gg J$. We remind that fluctuations of the single-particle levels are not taken into account in Eq.~\eqref{FinalNu}. 

As follows from Eq.~\eqref{FinalNu}, at small magnetic fields, $b\lesssim J$, the dependence of the TDOS on magnetic field is very weak. The TDOS for $b\lesssim J$ almost coincides with the result for $b=0$. Therefore, below we 
consider in detail the cases of zero ($b=0$) and large magnetic fields ($b\gg J$) only.

\begin{figure}[t]
\centerline{\includegraphics[width=75mm]{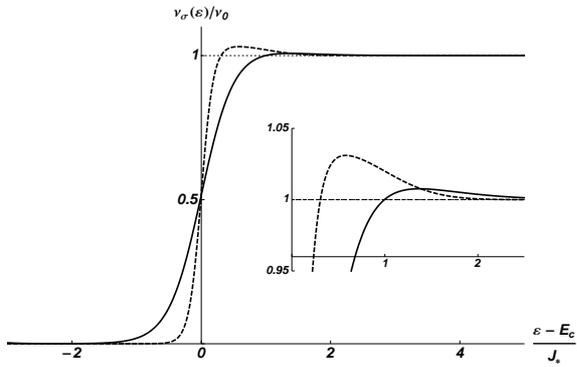}}
\caption{The tunneling density of states in the Coulomb valley. The solid
(dashed) line corresponds to $J/\delta =0.92$, $\delta/T=0.35$, and
$J_\star/T=3.95$ ($J/\delta =0.92$, $\delta/T=0.95$, and
$J_\star/T=10.70$). The inset depicts the nonmonotonic behavior.}
\label{FigureValley}
\end{figure}
%

\subsection{TDOS in zero magnetic field, $b=0$\label{NuZeroB}}

Expanding the expression in square brackets in the right hand side of Eq.~\eqref{FinalNu} to first order  on $b$, we find the following result for the TDOS in zero magnetic field:~\cite{BGK}
\begin{gather}
\frac{\nu_\sigma(\varepsilon)}{\nu_0} = \sum_{n, p=\pm}  e^{-\beta E_c(n-N_0)^2} \Biggl [
\left (1+\frac{J}{2J_\star}\right )
f_F(p \ve - 2p\Omega_n^{-p}) \notag \\
- \frac{J}{2J_\star}  \mathcal{F}\left (\frac{p \ve - 2 p \Omega_n^p}{J_\star},\beta J_\star\right )\Biggr ] \Biggl /\sum_n  e^{-\beta E_c(n-N_0)^2} .
\label{FinalNu0}
\end{gather}
Here the function $\mathcal{F}(x,y)$ is defined as
\begin{align}
\mathcal{F}(x,y) & = \frac{1}{2} e^{-y/4} e^{y x/2} \int_{-\infty}^\infty dt\, \frac{e^{i y x t -y t^2}}{\cosh(\pi t)}   . \label{mathFDef}
\end{align}
Using the following asymptotic expression at 
$y\ll 1$ for $\mathcal{F}(x,y)$: 
\begin{equation}
\mathcal{F}(x,y) = f_F(-y x) \left [ 1 - \frac{y}{2\cosh^2(yx/2)}\right ] ,
\end{equation}
we obtain the TDOS at $T\gg J_\star$:
\begin{align}
\frac{\nu_\sigma(\varepsilon)}{\nu_0} &= \sum_{n, p=\pm}  e^{-\beta E_c(n-N_0)^2}
f_F(p \ve - 2p\Omega_n^{-p})\Bigl \{ 1  \notag \\
+ & \frac{\beta J}{4\cosh^2(\frac{p}{2} \ve - p\Omega_n^{-p})}\Bigr \}
 \Biggl /\sum_n  e^{-\beta E_c(n-N_0)^2} . \label{eqHTd}
\end{align}
It is instructive to compare \eqref{eqHTd} with the result of Ref.~[\onlinecite{SeldmayrLY}] for the TDOS in the absence of exchange interaction. As expected, in this high temperature regime, the exchange interaction affects the tunneling density of states only slightly. It is worthwhile to mention that the correction  to the tunneling density of states due to exchange interaction is of the order of $J/T$ rather than what one may expect from above results on the spin susceptibility (see Eq.~\eqref{SpinSuscRegI_Fin}), namely, $J_\star/T$.

\begin{figure}[t]
\centerline{\includegraphics[width=75mm]{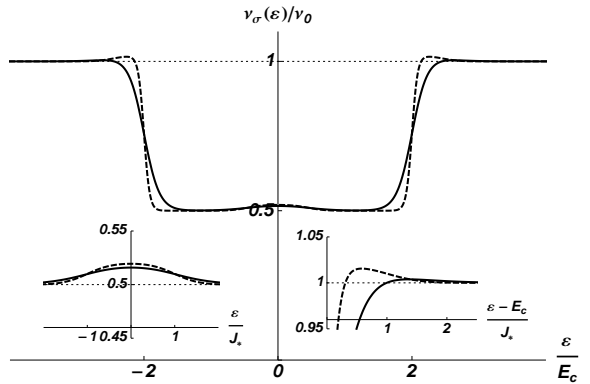}}
\caption{The tunneling density of states at the Coulomb peak.
The parameters are the same as in Fig.\protect\ref{FigureValley}.
The insets depict the nonmonotonic behavior.} \label{FigurePeak}
\end{figure}

In the regime of intermediate temperatures $\delta \ll T \ll J_\star$, it is convenient to use the
following result for the behavior of $\mathcal{F}(x,y)$ in the limit $y\gg 1$ (see Appendix~\ref{Appendix_MathF}):
\begin{align}
\mathcal{F}(x,y) & = \frac{1}{2}\sgn\left ( \cos\frac{\pi x}{2}\right ) e^{-\frac{y}{4}(x-1)^2 + \frac{y}{\pi^2} \cos^2 \frac{\pi x}{2}} \notag \\
\times& \left [ 1- \erf\left (\frac{\sqrt{y}}{\pi} \left |\cos\frac{\pi x}{2}\right |\right)\right ]+e^{\frac{y}{2}(x-|x|)} \notag\\
\times &\sum_{m\geqslant 0} (-1)^m e^{ - y |x| m + y m(m+1)} \theta(|x|-2m-1) . \label{tF_assympt}
\end{align}
Here $\theta(x)$ is the Heaviside step function ($\theta(0)\equiv 0$),
and the error function $\erf(z) = (2/\sqrt{\pi})\int_0^z
\exp(-t^2) dt$.

As $x$ is varied for a fixed $y$, Eq.~\eqref{tF_assympt} suggests
that $\mathcal{F}(x,y)$ exhibits damped oscillations with a period $4$
(equivalent to an energy scale $4J_\star$). However, it is not the case. At $y\gg 1$ the
function $\mathcal{F}(x,y)$ is monotonous and close to the function
$1/(1+\exp(-y(x-1)))$. The linear combination of two Fermi functions (standard one and one shifted in energy on $J_\star$) in Eq.~\eqref{FinalNu0} leads to the appearance of a maximum in the
TDOS. The height of the maximum can be approximately estimated as
$[\nu_\sigma(\varepsilon)/\nu_0]_{\rm max} -1 \sim J/J_\star$. This
additional structure in the TDOS reflects enhanced electron correlations
due to the exchange interaction.

In the case of temperatures $T\ll E_c$ we illustrate the non-monotonic behavior in the TDOS due to exchange interaction in Figs.~\ref{FigureValley} and \ref{FigurePeak} for the Coulomb valley ($N_0$ is integer) and
for the Coulomb peak ($N_0$ is half-integer).

%
\begin{figure}[t]
\centerline{\includegraphics[width=75mm]{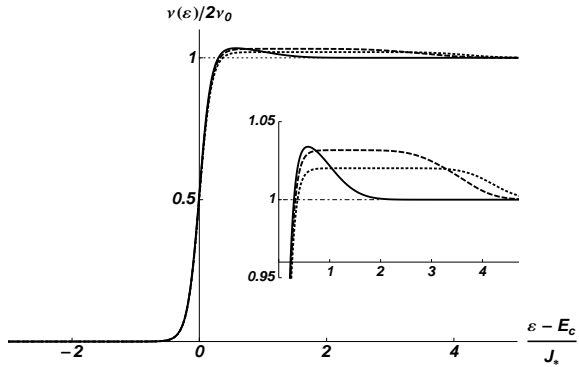}}
\caption{The total tunneling density of states $\nu(\ve)=\nu_\uparrow(\ve)+\nu_\downarrow(\ve)$ in the Coulomb valley for the magnetic field
$b=0$ (solid line), $b=2.75 J$ (dashed line) and $b=3.25J$ (dotted line). The other parameters are $J/\delta =0.92$, $\delta/T=0.95$, and
$J_\star/T=10.7$. The inset depicts the nonmonotonic behavior.}
\label{FigureValleyB}
\end{figure}
%

\subsection{TDOS at a strong magnetic field, $b\gg J$ \label{TDOS_LargeB}}

Now let us consider the case of a strong magnetic field, $b\gg J$. As above we consider the regime $\delta \ll T \ll J_\star$ near the Stoner instability, $J\ll J_\star$, in which the most interesting behavior of the TDOS takes place. 
To simplify the expression~\eqref{FinalNu} for the tunneling density of states, we use the following asymptotic result 
\begin{equation}
\mathbb{F}(x,y,z) \approx \frac{1}{zy\sqrt{\pi}} e^{z^2(y-1/2)^2} \mathcal{F}(2y-x/z^2,z^2) , \label{eq_asF}
\end{equation}
which is valid for $z\gg 1$ and $y \gg 1$. It is worthwhile to mention that Eq.~\eqref{eq_asF} works well already for $y\gtrsim 2$. With the help of Eq.~\eqref{eq_asF}, we obtain from Eq.~\eqref{FinalNu} the following expression for the TDOS
 \begin{align}
\frac{\nu_\sigma(\varepsilon)}{\nu_0} & = \sum_{n\in \mathbb{Z}} e^{-\beta E_c(n-N_0)^2}
\sum_{p=\pm}  \Biggl \{ f_F\bigl ( p\ve - 2 p \Omega_n^{-p}\bigr ) 
\notag \\
&+ \frac{J^2}{2J_\star (b-J)} 
\frac{e^{\sigma \beta b/4}}{\cosh(\beta b/4)}
\Bigl [ f_F\bigl ( p\ve - 2 p \Omega_n^{-p}+\sigma b/2\bigr ) \notag \\
&-
\mathcal{F}\bigl ((p\ve - 2 p \Omega_n^{p}-J_\star b/J)/J_\star, \beta J_\star\bigr )
\Bigr ] \Biggr \} \notag 
\\ & \times \Biggl [  \sum_{n\in \mathbb{Z}} e^{-\beta E_c(n-N_0)^2} \Biggr ]^{-1}.  \label{FinalNuBB}
\end{align}
As follows from Eq.~\eqref{FinalNuBB}, the non-monotonic behavior of the TDOS due to exchange interaction 
survives in the presence of magnetic field.  In Fig.~\ref{FigureValleyB} the dependence of the total TDOS $\nu(\ve)=\nu_\uparrow(\ve)+\nu_\downarrow(\ve)$ on energy at different magnetic fields 
is shown for the Coulomb valley. The role of magnetic field is of two kinds. At first, it suppresses the height of the maximum:
$[\nu(\varepsilon)/2\nu_0]_{\rm max} -1 \sim J\delta/(J_\star b)$. Secondly, the width of the maximum increases linearly ($\sim J_\star b/J$) with the magnetic field. Finally, we note that the difference $\nu_\uparrow(\ve)-\nu_\downarrow(\ve)$ is small.

\begin{figure}[t]
\centerline{\includegraphics[width=75mm]{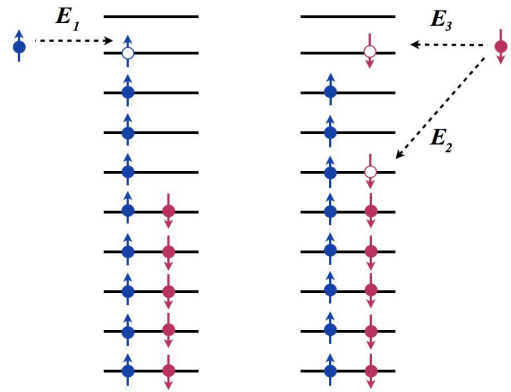}}
\caption{(Color online) Tunneling of an electron with spin up (left) and spin down (right) into a quantum dot with finite value of spin in the ground state
 (see text).} \label{FigureLevels}
\end{figure}

\section{The effect of level fluctuations on the TDOS in zero and strong magnetic fields \label{Sec_TDOS_Fluc}}

\begin{figure}[t]
\centerline{\includegraphics[width=75mm]{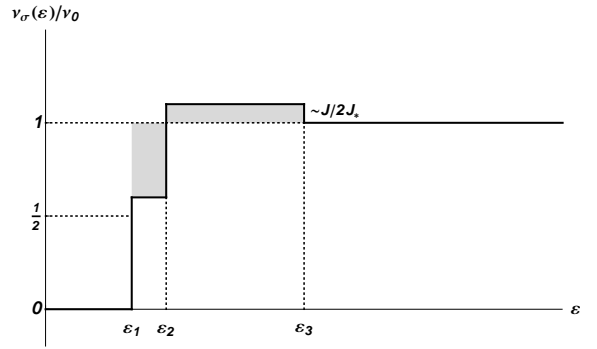}}
\caption{Sketch of dependence of the tunneling density of states on energy at zero temperature. Shaded areas are equal. (see text)} \label{FigureTDOS_Sketch}
\end{figure}

In order to develop qualitative explanation of the maximum in the TDOS and in order to qualitatively understand the effect of level fluctuations, it is instructive to consider a Coulomb valley at $T=0$. We start from the case of zero magnetic field. Let us consider tunneling of an electron with spin up to the ground state (with spin $S$) of a quantum dot in a Coulomb valley (see Fig.~\ref{FigureLevels}). The tunneling is possible only if the electron energy $\ve$ exceeds $\mathcal{E}_1 = E_{S+1/2}-E_S-J(S+3/4)$. Here we recall that $E_S$ denotes the single-particle contribution to the energy of the ground state with spin $S$. An electron with spin down can tunnel provided its energy is larger than $\mathcal{E}_2 =   E_{S-1/2}-E_S+J(S+1/4)$. Therefore, only tunneling of spin-up electrons is allowed in the energy interval $\mathcal{E}_1<\ve<\mathcal{E}_2$. At very large electron energies the tunneling is insensitive to the spin of electron. It is natural to denote the energy  $\mathcal{E}_3= E_{S+1/2}-E_S+J(S+1/4)$ as the characteristic energy above which there is no difference in tunneling for electrons with spin up and down. With the help of Eq.~\eqref{Crit2_0}, one finds $\mathcal{E}_2-\mathcal{E}_1\approx J-\Delta E_{2S-1}$ and $\mathcal{E}_3-\mathcal{E}_2\approx 2J S+\Delta E_{2S-1}$.  To sketch the tunneling density of states we employ the sum rule~\eqref{SumRule}. In the Coulomb valley and at zero temperature it renders the integral $\int d\ve \nu_\sigma(\ve)$ independent of $J$. The TDOS at $T=0$ is shown schematically in Fig.~\ref{FigureTDOS_Sketch}. The relative height of the maximum in the TDOS can be estimated as $(\mathcal{E}_2-\mathcal{E}_1)/(2(\mathcal{E}_3-\mathcal{E}_2)) \approx1/(4S)$. Near the Stoner instability, $J_\star\gg J$, the relative height becomes of the order of $J/(2J_\star)$ if one neglects the effect of level fluctuations. This estimate is in accordance with the result~\eqref{FinalNu0} derived for $\delta\ll T\ll J_\star$. In this temperature regime, the feature in the TDOS in the energy interval $\mathcal{E}_1< \ve< \mathcal{E}_2$ (see Fig.~\ref{FigureTDOS_Sketch}) is smeared, as expected, since $\mathcal{E}_2-\mathcal{E}_1\ll T$.

Our qualitative arguments can be justified by direct evaluation of the TDOS at zero temperature in a Coulomb valley. For the ground state of a quantum dot with spin $S$ we find from Eq.~\eqref{GK2} at zero magnetic field: 
\begin{align}
\nu_\sigma(\varepsilon) &= \frac{1}{2} \sum_{\epsilon_\alpha>\epsilon_{\frac{N_0}{2}-S}} \!\!\!\!
\delta\Bigl (\varepsilon-\epsilon_{\alpha} +\mu -E_c-J(S+1/4)\Bigr ) 
\notag \\
- & \frac{1}{4S+2} \sum_{\epsilon_\alpha>\epsilon_{\frac{N_0}{2}+S}}\!\!\!\!
\delta\Bigl (\varepsilon-\epsilon_{\alpha} +\mu -E_c-J(S+1/4)\Bigr ) 
\notag \\
+& \frac{S+1}{2S+1} \sum_{\epsilon_\alpha>\epsilon_{\frac{N_0}{2}+S}} \!\!\!\!
\delta\Bigl (\varepsilon-\epsilon_{\alpha} +\mu -E_c+J(S+3/4)\Bigr )  .
\end{align}
Using the fact that $\epsilon_{\frac{N_0}{2}+S+1}=E_{S+1/2}-E_S$ and $\epsilon_{\frac{N_0}{2}-S+1}=E_{S-1/2}-E_S$, we find
\begin{align}
\int_{\mathcal{E}_1}^{\mathcal{E}_2} d\ve\frac{\nu_\sigma(\varepsilon)}{\nu_0} & = \frac{1}{2}\left ( 1+\frac{1}{2S+1}\right ) (\mathcal{E}_2-\mathcal{E}_1) , \notag \\
\int_{\mathcal{E}_2}^{\mathcal{E}_3} d\ve\frac{\nu_\sigma(\varepsilon)}{\nu_0} & = \left ( 1+\frac{1}{4S+2} \right ) (\mathcal{E}_3-\mathcal{E}_2) \label{EqHeight}
\end{align}
in accordance with the sketch of Fig.~\ref{FigureTDOS_Sketch}.

We can estimate the effect of level fluctuations on the TDOS from the qualitative arguments presented above. As we have already demonstrated, at zero temperature the height of the maximum in the TDOS is given by $1/(4S)$ for a given realization of single-particle levels. Their statistical fluctuations lead to averaging of this estimated result, i.e., the height of the maximum in the average TDOS is given by $\overline{1/(4S)}\approx [1+\overline{(\Delta n_{2S})^2}] J/(2J_\star)$ (see Eq.~\eqref{Sfluc}). 
Taking into account Eq.~\eqref{Crit2} and the discussion after Eq.~\eqref{chi_fin_f_T0}, we expect that the height of the maximum in the average TDOS is of the order of  
$[1+(1/\bm{\beta}\pi^2)\ln(J_\star/T)]J/(2J_\star)$. Similarly to the zero-field average spin susceptibility the statistical fluctuations of single-particle levels result in logarithmic dependence with temperature of the height of the maximum in the TDOS.

In the presence of a strong magnetic field, $b\gg J$, and with neglect of level fluctutions, the the typical spin $S$ is of the order of $J_\star b/J\delta$. Therefore, the width of the maximum in the TDOS can be now estimated as $J S \sim J_\star b/\delta$. The relative height of the maximum becomes  $1/4S\sim J\delta/J_\star b$. Both estimates are in agreement with the results of Sec.~\ref{TDOS_LargeB}.

As was demonstrated in Sec.~\ref{RegIIISpinSusc}, a magnetic field, $b\gg J$, strongly suppresses the effect of level fluctuations on the spin susceptibility. The same holds for the TDOS. The level fluctuations result in relative corrections to the TDOS~\eqref{FinalNuBB} of the order of  $(\delta/b) \ln (J_\star b/J\delta) \ll 1$.

\section{Conclusions\label{Sec.Conc}}

We have addressed here the interplay of charging and spin-exchange interactions of electrons 
in a quantum dot. Even within the simple UH framework, this problem becomes non-trivial due to the underlying non-Abelian action that necessarily requires the evaluation of time-ordered integrals. To overcome this obstacle we have employed here a technique based on the WNK transformation. It allows us to obtain exact analytic resuts which describe the partition function (Eq.~\eqref{GCPF_Gen}) and the tunneling density of states (Eq.~\eqref{GK2}) for an arbitrary single-particle spectrum, temperature, Zeeman splitting, charging and exchange energies. Our solution~\eqref{GCPF_Gen} for the partition function reproduces the result obtained previously by means of another approach in Refs.~[\onlinecite{AlhassidRupp,AlhassidHernando}]. We believe that the approach employed in this paper is more manageable for analytic calculations and extensions. Our result~\eqref{GK2} for the TDOS is a generalization of the result derived in Ref.~[\onlinecite{SeldmayrLY}] to the case of finite spin-exchange interaction and Zeeman splitting.

In the mesoscopic Stoner regime, near the Stoner instability, $\delta-J\ll \delta$, we have analyzed our general results~\eqref{GCPF_Gen} and ~\eqref{GK2} in detail. In particular, we have found that in a wide temperature range $\delta \ll T \ll \delta J/(\delta-J)$, the average zero-field spin susceptibility behaves according to the Curie law with a large effective spin which depends  on temperature logarithmically (see Eq.~\eqref{chi_fin_f}). This dependence results from statistical fluctuations of single-particle levels in QDs. We have demonstrated that a tiny magnetic field $B\sim \sqrt{JT(1-J/\delta)}/g\mu_B$ is sufficient to suppress temperature dependence of the average spin susceptibility, as well as to diminish a role of statistical fluctuations. We have found that enhanced spin correlations, resulting in a large total spin in the ground state of a QD (in the mesoscopic Stoner regime, $\delta-J\ll \delta$), become apparent as additional (to Coulomb blockade) non-monotonic behavior in the TDOS (at high temperatures, $\delta \ll T \ll \delta J/(\delta-J)$). Similarly to the case of the spin susceptibility, we have found that magnetic field suppresses the spin-related non-monotonic behavior of the TDOS.

To test our results one needs to explore QDs made of materials close to the thermodynamic Stoner instability.  The long list of such materials includes  Co impurities in a Pd or Pt host, Fe or Mn dissolved in various
transition-metal alloys, Ni impurities in a Pd host, and Co in Fe grains, as well as new, nearly ferromagnetic rare-earth
materials.~\cite{Exp1,Exp2,Exp3} For the closest material to the Stoner instability, we are aware of, YFe$_2$Zn$_{20}$, the parameter $J/(\delta-J)$ is approximately equal to $16$, leading to the spin in the ground state of the order of $10$. Our results for the spin susceptibility can be checked by measuring of the total magnetization and the electronic spin resonance signal. Although, as is well-known (see e.g. Ref.~[\onlinecite{Korenblit}]), a single Fe impurity in a nearly ferromagnetic material typically acquires  an effective spin, $S_{\rm imp}$, of the order of $10$, the temperature behavior of the impurity's contribution to the spin susceptibility is different: $\chi_{\rm imp} \sim  S^2_{\rm imp}/(T \ln T/T_0)$, where the temperature $T_0$ is set by  the interactions between the electron spins and the impurity.~\cite{LarkinMelnikov} We speculate that the intriguing magnetic behavior 
observed recently in Pd nanoparticles capped with different protective systems~\cite{Exp4} can be related to the physics of mesoscopic Stoner regime. Our predictions for the TDOS can be manisfested in the non-linear current-voltage characteristics of QDs,
measured in the sequential tunneling regime at temperatures lower than the bias voltage. The expected spin-related non-monotonicities may exceed $~7-15\%$ and should be sensitive to the applied magnetic field.

The approach used here for analytic computation of the spin susceptibility and the TDOS in the UH with isotropic spin-exchange interaction can be applied for other problems. For example, it can be used for the study of transport through a QD described by the UH in the co-tunneling regime that requires calculation of two-particle Green's functions of electrons in QD. One more example is provided by the UH with anisotropic (XXZ) spin-exchange interaction.~\cite{UsajBaranger2} This model is of interest since it describes a crossover from the case of isotropic spin-exchange interaction with mesoscopic Stoner instability and trivial dynamical spin susceptibility to the case of Ising spin-exchange interaction without mesoscopic Stoner regime but with interesting dynamical spin susceptibility.  Also our approach can be fruitful for investigation of the same crossover from isotropic to Ising spin-exchange interaction realized in QDs with spin-orbit interaction.~\cite{AlhassidSO,Murthy,AlhassidTureci} Yet another example is given by the competition between superconductivity and ferromagnetism in the UH with an attractive interaction in the Cooper channel.~\cite{AlhassidSF} The method used in the present paper can be generalized to study this interplay as well.


\begin{acknowledgements}

We acknowledge useful discussions with I. Aleiner, A. Altland, V. Glazkov, V. Gritsev, M. Feigelman, V. Kravtsov, Yu. Makhlin, G. Murthy, A. Shnirman, M. Skvortsov, and G. Tsirlina. We thank Y. Alhassid for explaining to us his method and the results of
his analysis. We are grateful to I. Kolokolov for providing us with notes of his calculations and a detailed explanation. The research was funded in part by the Russian-Israel scientific research cooperation (RFBR Grant No. 11-02-92470 and IMOST  3-8364), the Council for Grant of the President of Russian Federation (Grant No. MK-296.2011.2), RAS Programs ``Quantum Physics of Condensed Matter'' and ``Fundamentals of nanotechnology and nanomaterials'', the Russian Ministry of Education and Science under contract No. P926, the Minerva Foundation, the German-Israel GIF, the Israel Science Foundation of the Israel Academy of Sciences and Humanities, and EU project GEOMDISS.

\end{acknowledgements}


\appendix

\section{Wei-Norman-Kolokolov transformation\label{Appendix_WNK}}

In this appendix we demonstrate relation of the non-linear transformation~\eqref{VarChg} with the general method by Wei and Norman~\cite{WeiNorman} and calculate the Jacobian \eqref{Jac_WNK} of Wei-Norman-Kolokolov transformation.~\cite{Kolokolov}

The equation for time evolution operator for the time dependent
Hamiltonian $H(t)$
\begin{eqnarray}\label{wnk1}
i\frac{d U}{d t}= H(t) U(t)
\end{eqnarray}
with initial condition $U(0)=1$ can be represented by a finite
product of $k$ exponential operators.~\cite{WeiNorman} The index $k$
is a dimension of the Lie algebra generated by $H(t)$ where the
Hamiltonian is assumed to be linearly dependent on the group
generators. The general formalism of time evolution operator
construction is known as Wei-Norman method.~\cite{WeiNorman} 

Consider a time-evolution operator for the system with linear
realization of dynamical $SU(2)$ symmetry described by the
Hamiltonian
\begin{eqnarray}\label{wnk2}
\hat H(t)=-\bm{\theta}(t)  \bm{s}
\end{eqnarray}
Then, matrix $U(t)$ is given by time-ordering exponent~\eqref{eq_A}.
The solution of (\ref{wnk1}) can be parameterized as
\begin{eqnarray}\label{wnk3}
U(t)=\exp(f_-(t)s^-)\exp(f_z(t) s^z)\exp(f_+(t)s^+)
\end{eqnarray}
where $s^\pm =s^x\pm i s^y$ and $s^z$ are three generators of
$SU(2)$ group. Functions $f_-(t)$, $f_z(t)$ and $f_+(t)$ satisfy the
system of differential equations
\begin{eqnarray}\label{wnk4}
\left\{
\begin{array}{ccc}
i \dot f_-&=&-\frac{1}{2}\theta^++\theta^z f_- +\frac{1}{2}\theta^- f_-^2\\
i \dot f_z&=&-\theta^z-\theta^- f_-\\
i \dot f_+&=&-\frac{1}{2}\theta^-e^{-f_z}
\end{array}\right.
\end{eqnarray}
with initial condition $f_-(0)=f_z(0)=f_+(0)=0$. This system of equations can be easily
obtained with the help of Hausdorf formula.~ \cite{WeiNorman} The
solution of the system (\ref{wnk4}) depends on the solution of the
single Riccati equation (the first equation in (\ref{wnk4})).
Parametrization of (\ref{wnk4}) by three new functions
$\kappa^\pm(t), \rho(t)$ defined as
\begin{align}
\kappa^+(t)&=f_-(t),\notag \\
\kappa^-(t) &=-i\dot f_+(t)
e^{f_z(t)},\label{wnk5} \\
\rho(t) &=-i\dot f_z(t).
\end{align}
leads to  the Kolokolov representation of the
time-evolution operator:~\cite{Kolokolov}
\begin{align}
U(t) &=e^{\kappa^+(t) s^-} e^{i s^z\int_0^t\rho(t_1)
dt_1}\notag \\& \times \exp\left(is^+\int_0^t\kappa^-(t_1)e^{-i\int_0^{t_1}\rho(t_2)
dt_2}dt_1\right) . \label{wnk6}
\end{align}
The initial condition $f_-(0)=0$ is translated to $\kappa^+(0)=0$. The
initial conditions $f_z(0)=f_+(0)=0$ are satisfied by construction of
the functions $\kappa^-(t)$ and $\rho(t)$. Therefore, the variables $\kappa^-(t)$ and $\rho(t)$ are not constrained by the initial conditions.  

In order to find the Jacobian of the Wei-Norman-Kolokolov transformation, we consider its discrete version. Let us split the interval $(0,t)$ into $N\to\infty$ parts of length $\Delta\to 0$ such that $t = N \Delta$. Also we introduce $\bm{\theta}_j\equiv \bm{\theta}(t_j)$, $\kappa^{\pm}_j \equiv \kappa^\pm(t_j)$ and $\rho_j\equiv\rho(t_j)$, where $t_j = j \Delta$ with  $j=0,\dots,N$. Then, 
the solution of Eqs~\eqref{wnk1}-\eqref{wnk2} can be written as $U(t) = \prod_{j=0}^N \exp(i\Delta \bm{\theta}_j \bm{s})$. Together with Eq.~\eqref{wnk6} it allows us to find the following map corresponding to Wei-Norman-Kolokolov transformation (see Eq.~\eqref{VarChg} with $p=+$):
\begin{align}
\theta_j^- &= \kappa_j^-  , \notag \\
\theta_j^+ &= -i\frac{\kappa_j^+-\kappa_{j-1}^+}{\Delta} +\rho_j\frac{\kappa_j^++\kappa_{j-1}^+}{2} - \kappa_{j-1}^+\kappa_j^-\kappa_j^+ ,\label{wnk7}\\
\theta_j^z & = \rho_j-\kappa_j^-(\kappa_j^++\kappa_{j-1}^+) . \notag 
\end{align}
The map~\eqref{wnk7} is supplemented by the initial condition $\kappa_0^+=0$. The Jacobian of the map~\eqref{wnk7} is given as
\begin{equation}
\mathcal{J}_N = \prod_{j=1}^N \left (-\frac{i}{\Delta}+\frac{\rho_j +(\kappa_j^+-\kappa_{j-1}^+)\kappa_j^-}{2}\right ) .
\end{equation}
Finally, taking the limit $N\to \infty$ and $\Delta\to 0$, we obtain~\cite{Kolokolov}
\begin{equation}
\mathcal{J} = \left (i\Delta\right )^{-N} \exp \left ( \frac{i}{2}\int_0^t dt^\prime \rho(t^\prime) \right ) 
\end{equation}
in agreement with Eq.~\eqref{Jac_WNK}.

\section{Integration over $\kappa_p^{\pm p}$ in Eq.~\eqref{ZK3}\label{Appendix_IntWNK}}

In this appendix we demonstrate how integration in Eq.~\eqref{ZK3} over fields $\kappa_p^{\pm p}$ can be performed. 
We need to evaluate the following functional integral: 
\begin{gather}
\left [ \prod_{p=\pm}
\int \mathcal{D}[\kappa_{p}^{\pm p}] e^{-\int_0^{t_p}dt \dot{\kappa}_p^p \kappa_p^{-p}/J} \right ]
\exp \Biggl \{ v \prod_{p=\pm}e^{\frac{i p}{2} \int\limits_0^{t_p}dt \tilde{\rho}_p(t)} \notag \\
\times \Bigl [ i p\tilde{\kappa}_p^p(t_p)- \int\limits_0^{t_{-p}}dt\,  \tilde{\kappa}_{-p}^p(t) e^{i p \int\limits_0^t dt^\prime \tilde{\rho}_{-p}(t^\prime)}\Bigr ]  \Biggr \} . \label{CorFB0}
\end{gather}

Let us start from computation of  the 2-point correlation function
\begin{equation}
\langle \kappa_p^p(t_p) \kappa_p^{-p}(t)\rangle_0 \equiv \frac{\int \mathcal{D}[\kappa_{p}^{\pm p}]  \kappa_p^p(t_p) \kappa_p^{-p}(t) e^{-\int_0^{t_p}dt \dot{\kappa}_p^p \kappa_p^{-p}/J}} 
{\int \mathcal{D}[\kappa_{p}^{\pm p}] e^{-\int_0^{t_p}dt \dot{\kappa}_p^p \kappa_p^{-p}/J}} .\label{CorFB1}
\end{equation}
In order to evaluate the functional integral, we split the interval $(0,t_p)$ into $N$ parts of length $\Delta\to 0$ such that $t_p = N \Delta$. We define $\kappa^{\pm p}_{p,j} \equiv \kappa^{\pm p}_p(t_j)$, where $t_j = j \Delta$ with  $j=0,\dots,N$. Then, the 2-point correlation function \eqref{CorFB1} can be written as ($t=n\Delta$):
\begin{equation}
\langle \kappa^p_{p,N} \kappa^{-p}_{p,n}\rangle_0 = 
\frac{\bigl(\prod_{j=1}^N \int d\kappa_{p,j}^{p}d\kappa_{p,j}^{-p}]\bigr)  \kappa_{p,N}^p\kappa_{p,n}^{-p} e^{-\bm{\kappa}_p^p \bm{\Gamma}^{-1} \bm{\kappa}_p^{-p}}} 
{\bigl(\prod_{j=1}^N \int d\kappa_{p,j}^{p}d\kappa_{p,j}^{-p}]\bigr) e^{-\bm{\kappa}_p^p \bm{\Gamma}^{-1} \bm{\kappa}_p^{-p}}}  . \label{CorFB2}
\end{equation}
Here $\bm{\kappa}_p^{\pm p} = (\kappa_{p,0}^{\pm p},\dots,\kappa_{p,N}^{\pm p})$ and  
\begin{equation}
\bm{\Gamma} =J  \begin{pmatrix}
1&1&1&\dots &1\\
0&1&1&\dots&1\\
\dots &\dots & \dots & \dots & \dots \\
0&0&0&\dots &1
\end{pmatrix} .
\end{equation}
We remind that $\kappa_{p}^p$ satisfies the initial condition $\kappa_{p,0}^p=0$. Hence, we find
\begin{equation}
\langle \kappa_p^p(t_p) \kappa_p^{-p}(t)\rangle_0 = J \theta(t_p-t), 
\end{equation}
where the Heaviside step function is defined as $\theta(0)=1$. For obvious reasons, the other 2-point correlation functions vanish.
Next, we can find the following set of identities for an arbitrary function $f(t)$ and non-negative integer $k$:
\begin{gather}
\left \langle \left ( \kappa_p^p(t_p) \int_0^{t_p} dt\, f(t) \kappa_p^{-p}(t)\right )^k \right \rangle_0
 = k! \left (J  \int_0^{t_p} dt\, f(t)\right )^k \notag \\
= \frac{\int d\varkappa_pd\varkappa_p^* \,e^{- |\varkappa_p|^2/J} \bigl ( \varkappa_p\int_0^{t_p} dt\, f(t) \varkappa_p^*\bigr )^k }{\int d\varkappa_pd\varkappa_p^*\, e^{- |\varkappa_p|^2/J}} .
\end{gather}
We see that the functional integral \eqref{CorFB0} can be substituted by the usual one with complex conjugated variables $\varkappa_p$ and $\varkappa_p^*$ corresponding to $\kappa_p^{p}$ and $\kappa_p^{-p}$, respectively. Therefore, Eq.~\eqref{CorFB0} becomes
\begin{gather}
\left [ \prod_{p=\pm}
\int \frac{d\varkappa_{p}d\varkappa_{p}^*}{\pi} e^{-|\varkappa_p|^2/J} \right ]
\exp \Biggl \{ v \prod_{p=\pm}e^{\frac{i p}{2} \int\limits_0^{t_p}dt \tilde{\rho}_p(t)} \notag \\
\times \Bigl [ i p e^{i b t_p}\varkappa_p- \varkappa_{-p}^* \int\limits_0^{t_{-p}}dt\,  e^{i p \int\limits_0^t dt^\prime \rho_{-p}(t^\prime)}\Bigr ]  \Biggr \} . \label{CorFB3}
\end{gather}
Integrating over variables $\varkappa_p$ and $\varkappa_p^*$, we find that integral~\eqref{CorFB3} equals
\begin{gather}
J^2\Biggl [ 1+ i J v \left ( \prod_{p=\pm} e^{\frac{ip}{2} \int_0^{t_p}dt \rho_p} \right )\Biggl (\sum\limits_{p=\pm} p e^{\frac{i p b}{2}(t_+-t_-)}
\notag \\
\times \int_0^{t_p} dt\, e^{-i p \int_0^t dt^\prime \rho_p(t^\prime)} \Biggr ) \Biggr ]^{-1} .
\end{gather}
This result yields Eq.~\eqref{ZK4}.


\section{Correlation function $C(h_1,h_2)$ \label{App_CorrFunc}}

In this appendix we present derivation of the results~\eqref{Sassymp} for the correlation function~\eqref{DefCorr}. Also, we derive the result~\eqref{FlucRes}. 

The single-particle density of states $\nu_0(E)$ has non-Gaussian statistics. However, the function $V(h)$ is a Gaussian random variable~\cite{Mehta} since it involves a large number of single-particle levels: $\max\{|h|,T\}/\delta\gg 1$. We remind that the 2-point correlation function of the single-particle density of states is given as~\cite{Mehta}
\begin{equation}
\langle \delta\nu_0(E)\delta\nu_0(E+\omega)\rangle = \frac{1}{\delta^2} \left [\delta\left (\frac{\omega}{\delta}\right )- R_{U/O}\left (\frac{\pi \omega}{\delta}\right )\right ] ,
\label{App_dnu1}
\end{equation}
where 
\begin{align}
R_U(x) &= \frac{\sin^2x}{x^2} ,\\ 
R_O(x) &= \frac{\sin^2x}{x^2} +\left( \frac{d}{dx}\frac{\sin x}{x}\right ) \int_x^\infty \frac{\sin t}{t} dt  .
\end{align}
Using Eqs~\eqref{Vh_Def}, \eqref{DefCorr_0} and \eqref{App_dnu1}, we obtain 
\begin{align}
C(h_1,h_2) & = T^2 \int \frac{dE d\omega}{\delta^2} R_{U/O}\left (\frac{\pi T \omega}{\delta}\right )\notag\\
&\times \Biggl \{
\ln \left [ 1+\frac{\sinh^2(\frac{h_1}{2})}{\cosh^2(\frac{E}{2})}\right ] \ln \left [ 1+\frac{\sinh^2(\frac{h_2}{2})}{\cosh^2(\frac{E}{2})}\right ] \notag \\
&-
\ln \left [ 1+\frac{\sinh^2(\frac{h_1}{2})}{\cosh^2(\frac{E+\omega/2}{2})}\right ] \notag \\
&\times \ln \left [ 1+\frac{\sinh^2(\frac{h_2}{2})}{\cosh^2(\frac{E-\omega/2}{2})}\right ]
 \Biggr \} ,\label{eqC_app}
\end{align}
where we used the identity 
\begin{equation}
\int_{-\infty}^\infty R_{U/O}(x) dx=\pi.
\end{equation}
 Let us introduce the function 
 \begin{equation}
 C_{nm}(h_1,h_2) = \frac{d^{n+m}C(h_1,h_2)}{d^nh_1d^mh_2}
 \end{equation}
 with integers $n, m \geqslant 1$. 
Then one can check that the following exact relation holds
\begin{equation}
C_{11}(h_1,h_2) = L_2(h_1+h_2)-L_2(h_1-h_2) .\label{eqL8_app}
\end{equation}
Here, the function $L_2(h)$ is given as
\begin{align}
L_2(h)  &= 16T^2\sinh^2\frac{h}{2}  \int_0^\infty \frac{d\omega}{\delta^2} R_{U/O}\left (\frac{2\pi T \omega}{\delta}\right )\notag\\
 &\times \Biggl [ \frac{\frac{h}{2}\coth\frac{h}{2} -1}{\cosh h-1} -\frac{\frac{h}{2}\coth\frac{h}{2}- \omega\coth\omega}{\cosh h -\cosh 2\omega}\Biggr ] .\label{eqL2_app}
\end{align}
Using the conditions $C(h_1,h_2)=C(h_2,h_1)$ and $C(h,0)=0$, we obtain Eq.~\eqref{DefCorr} in which the function $L(h)$ is related with $L_2(h)$
as
\begin{equation}
L_2(h) = L^{\prime\prime}(h) . \label{eqL31_app}
\end{equation}

To estimate the function $L_2(h)$ at $T\gg\delta$, we can use the asymptotic expression of the function $R_{U/O}$ at large values of its argument: 
\begin{equation}
R_{U/O}(x) = \frac{1}{\bm{\beta} x^2} , \qquad x\gg 1 .
\end{equation}
Here we recall that $\bm{\beta}=1$ for the orthogonal ensemble and $\bm{\beta}=2$ for the unitary ensemble. Then, at $T\gg\delta$ Eq.~\eqref{eqL2_app} 
becomes
\begin{align}
L_2(h) & = \frac{4\sinh^2(h/2)}{\bm{\beta}\pi^2}  \int_0^\infty \frac{d\omega}{\omega^2} \Biggl [ \frac{\frac{h}{2}\coth\frac{h}{2} -1}{\cosh h-1} \notag \\
& -\frac{\frac{h}{2}\coth\frac{h}{2}- \omega\coth\omega}{\cosh h -\cosh 2\omega}\Biggr ] .\label{eqL3_app}
\end{align}

At $|h|\ll 1$ we expand the right hand side of Eq.~\eqref{eqL3_app} and find
\begin{equation}
 L_{2}(h) =\frac{c_1 h^2}{2\bm{\beta} \pi^2} , \label{eqL5_app}
\end{equation}
where the numerical constant
\begin{equation}
c_1 =  \int_0^\infty \frac{d\omega}{\omega^2} \Biggl \{
\frac{1}{3} - \frac{\omega\coth\omega -1}{\sinh^2\omega} 
 \Biggr \} \approx 0.37 .
\end{equation}
Hence, using Eq.~\eqref{eqL31_app}, we obtain the asymptotic expression~\eqref{Sassymp}.

At $|h| \gg 1$, we rewrite Eq.~\eqref{eqL3_app} as
\begin{align}
 L_{2}(h) &=\frac{2\sinh^2(h/2)}{\bm{\beta} \pi^2}  \Biggl \{
 \notag \\
 & \times 
 \int_0^1 \frac{d\omega}{\omega^2} \Bigl [ \frac{|h|-2}{2\sinh^2(h/2)} - \frac{|h| - 2\omega \coth\omega}{\cosh (h)}\Bigr ] \notag \\
 &
 +  \int_1^{|h|/2} \frac{d\omega}{\omega^2} \Bigl [ \frac{|h|-2}{2\sinh^2(h/2)} - 2 \frac{|h| - 2\omega \coth\omega}{e^{|h|}-e^{2\omega}}\Bigr ] \notag \\ 
& + \int_{|h|/2}^\infty \frac{d\omega}{\omega^2} \Bigl [ \frac{|h|-2}{2\sinh^2(h/2)} - 2 \frac{|h| - 2\omega}{e^{|h|}-e^{2\omega}}\Bigr ]
 \Biggr \} \notag \\
 &\approx  \frac{2}{\bm{\beta} \pi^2}  \left ( \ln \frac{|h|}{2} +c_2 \right ) ,\label{eqL4_app}
 \end{align}
 where
 \begin{equation}
  c_2 = -\int_0^1\frac{d\omega}{\omega^2}[1-\omega\coth\omega]+\int_1^\infty \frac{d\omega\,\ln\omega}{\sinh^2\omega} \approx 0.43 .
 \end{equation}
Using Eq.~\eqref{eqL31_app}, we obtain the asymptotic expression~\eqref{Sassymp} from Eq.~\eqref{eqL4_app}.

Equation~\eqref{Vh_small} implies that mean squared fluctuations of the level spacing $\Delta$ can be written as
\begin{equation}
\frac{\overline{(\Delta-\delta)^2}}{\delta^2} = C_{22}(0,0)\frac{\delta^2}{4T^2}  . \label{eqL7_app}
\end{equation}
As follows from Eqs~\eqref{eqL8_app} and \eqref{eqL5_app}, $C_{22}(0,0) = 2c_1/\bm{\beta}\pi^2$. Hence, we obtain Eq.~\eqref{FlucRes} from Eq.~\eqref{eqL7_app}.

\section{Evaluation of $\chi(T,0)$ in the region II$_a$ \label{Appendix_Replica}}

In this appendix we perform evaluation of the averaged zero-field spin susceptibility in the leading logarithmic approximation: in each order of expansion of $\overline{\chi(T,0)}$ in powers of $1/(\bm{\beta}\pi^2)$ we take into account the term with the highest power of $\ln 2J_\star/T$. 

Let us define $\Xi_{10}(x,y) = \partial \Xi(x,y)/\partial x$, then at $\delta\ll  T\ll J_\star$ ($y\gg 1$)
\begin{equation}
\overline{\chi(T,0)} = \frac{(J_\star/J)^2}{3 T} \frac{d}{dy}\overline{\ln \Xi_{10}(0,y)} . \label{EqE1}
\end{equation}
In order to evaluate $\overline{\ln \Xi_{10}(0,y)}$ we use the replica trick. For non-negative integer values of $n$ we obtain that 
\begin{align}
\overline{\bigl[\Xi_{10}(0,y)\bigr]^n} &=  \prod_{j=1}^n\left  [ y^{3/2} e^{y/4}\int \frac{du_j}{2} e^{-u_j^2} 
\bigl ( 1+ \frac{2u_j}{\sqrt{y}}\bigr )\right ] \notag \\
&\hspace{-1cm} \times \exp\left [\frac{1}{2}\sum_{j,k=1}^n C\Bigl(\frac{y}{2}+\sqrt{y} u_j, \frac{y}{2}+\sqrt{y} u_k\Bigr)\right ] . \label{eqCorrF2}
\end{align}
Provided the dominant contribution to the integral \eqref{eqCorrF2} comes from regions with $|u_j| \ll \sqrt{y}$, we can expand the 2-point correlation function~\eqref{DefCorr} in powers of $u_j$ and $u_k$ to the second order whereever it possible. We thus find 
\begin{align}
C\Bigl(\frac{y}{2}+\sqrt{y} u_j, \frac{y}{2}+\sqrt{y} u_k\Bigr)& \approx 
\frac{y^2\ln 2}{\bm{\beta}\pi^2}+\frac{1-a}{n}(u_j^2+u_k^2)  \notag \\
& \hspace{-1.5cm}  + 2 b u_j u_k + \frac{c}{n} (u_j+u_k) 
\notag \\ & \hspace{-1.5cm} + \frac{y}{2\bm{\beta}\pi^2} (u_j-u_k)^2\ln(u_j-u_k)^2, \label{CorAppC}
\end{align}
where we introduce the following parameters for convenience:
\begin{align}
a &=1+\frac{n y}{2\bm{\beta}\pi^2}\Bigl ( \ln \frac{y}{16}+3\Bigr ), \notag\\
b & = \frac{y}{2\bm{\beta}\pi^2}\Bigl ( \ln y+3\Bigr ) \notag \\
c & = \frac{2 n y^{3/2}}{\bm{\beta}\pi^2}\ln 2 .
\end{align}
The last term in the right hand side of Eq.~\eqref{CorAppC} can be described as the contribution from independent Gaussian variable
$v(u)=v(-u)$ with the correlation function
\begin{equation}
\overline{v(u_j)v(u_k)} = \frac{y}{2\bm{\beta}\pi^2} (u_j-u_k)^2\ln(u_j-u_k)^2 . \label{CorAppv}
\end{equation}
Then, we find
\begin{align}
\overline{\bigl[\Xi_{10}(0,y)\bigr]^n} &=  \prod_{j=1}^n\left  [ y^{3/2} e^{y/4}\int \frac{du_j}{2} e^{-a u_j^2+c u_j} 
\Bigl ( 1+ \frac{2u_j}{\sqrt{y}}\Bigr )\right ] \notag \\
&\hspace{-1cm} \times \exp\left [\frac{n^2 y^2 \ln 2}{2\bm{\beta}\pi^2}+b \left ( \sum_{j=1}^n u_j\right )^2\right ] \overline{\prod_{j=1}^n e^{v(u_j)}} . \label{AppCorE}
\end{align} 
Here we should still perform averaging over $v(u)$ with the help of Eq.~\eqref{CorAppv}. Introducing Gaussian variable $z$ to decouple term $\left ( \sum_{j=1}^n u_j\right )^2$ in the right hand side of Eq.~\eqref{AppCorE}, we rewrite it as
\begin{align}
\overline{\bigl[\Xi_{10}(0,y)\bigr]^n} &=  y^{3n/2} e^{ny/4}\exp\left [\frac{n^2 y^2 \ln 2}{2\bm{\beta}\pi^2}\right ]
\int \frac{dz}{\sqrt\pi} e^{-z^2}   \notag \\
&\hspace{-1.5cm} \times \prod_{j=1}^n\left  [\int \frac{du_j}{2} e^{-a u_j^2+(c+2z\sqrt{b}) u_j} 
\Bigl ( 1+ \frac{2u_j}{\sqrt{y}}\Bigr )\right ] \overline{\prod_{j=1}^n e^{v(u_j)}} . \label{AppCorE0}
\end{align}
We note that the typical values of $u_j$ contributing to the integral in Eq.~\eqref{AppCorE0} is of the order of $(c+2 z\sqrt{b})/a$. Next we introduce the variables $x_j=u_j-(c+2 z\sqrt{b})/2a$. Provided typical values of $|c+2 z\sqrt{b}|/2a\ll \sqrt{y}$, the  limits of integration over $x_j$ are the same as for $u_j$. Taking into account that the correlation function \eqref{CorAppv} is translationally invariant, we thus find  
\begin{align}
\overline{\bigl[\Xi_{10}(0,y)\bigr]^n} &=  
\frac{\pi^{n/2} y^{3n/2} e^{n (y +c^2/(a-nb))/4}}{2^n a^{(n-1)/2} (a-n b)^{1/2}}
e^{\frac{n^2 y^2 \ln 2}{2\bm{\beta}\pi^2}}
 \notag \\
&\times  
\int \frac{dz}{\sqrt\pi} e^{-z^2} \Bigl (1-\frac{2 z \sqrt{b(a-nb)}+c\sqrt{a}}{(a-nb)\sqrt{a y}}\Bigr )^n
 \notag \\
&\times  
 \overline{\left (\int \frac{dx}{\sqrt\pi}  e^{-x^2+v(x/\sqrt{a})} \right )^n }.
\end{align}
Here, we used the fact that $v(x)$ is an even function. In the limit $n\to 0$, we obtain 
\begin{align}
\overline{\ln\Xi_{10}(0,y)} & = \frac{y}{4} \left [1+\frac{1}{\bm{\beta}\pi^2} \Bigl (\ln y+ 3\Bigr )\right ] +\overline{\ln \mathcal{X}} \notag \\
&\hspace{-.5cm}+\int \frac{dz}{\sqrt{\pi}} e^{-z^2} \ln\Bigl [ 1+z\sqrt{2\ln y/(\bm{\beta}\pi^2)} \Bigr ]  ,\label{EqCorrF1}
\end{align}
where 
\begin{equation}
\mathcal{X} = \int \frac{du}{\sqrt\pi} e^{-u^2+v(u)} .
\end{equation}
As we discussed above, the integral over $z$ in Eq.~\eqref{EqCorrF1} is constrained by the condition $|z| \sqrt{\ln y/(\bm{\beta}\pi^2)}  \ll 1$. In this case, the typical values of $z$ contributing to the integral are of the order of unity. Our assumption is thus self-consistent for 
$ \sqrt{\ln y/(\bm{\beta}\pi^2)} \ll 1$. Therefore, the integral over $z$ in Eq.~\eqref{EqCorrF1} is  proportional to $\ln y/\bm{\beta}\pi^2\ll 1$, i.e., the integral is small compared to the first term.

Evaluating $\overline{\ln \mathcal{X}}$ to the lowest order in $y/(\bm{\beta}\pi^2)$ with the help of Eq.~\eqref{CorAppv} we find
\begin{align}
\overline{\ln \mathcal{X}}  = & a_1 \frac{y}{\bm{\beta}\pi^2} +\dots , \\
 & a_1  = (\ln 2+\gamma-2)/4\approx -0.18 .
\end{align}
Subtituting this result into Eq.~\eqref{EqCorrF1}, we reproduce Eq.~\eqref{chi_fin_f} with the help of Eq.~\eqref{EqE1}. 
To the second order in $y/(\bm{\beta}\pi^2)$  we find 
\begin{equation}
\overline{\ln \mathcal{X}}  =  a_1 \frac{y}{\bm{\beta}\pi^2} + a_2 \left (\frac{y}{\bm{\beta}\pi^2}\right )^2 + \dots , 
\label{a2S}
\end{equation}
where 
\begin{align}
a_2 & =  -\frac{3}{8} \left (\ln 2 + \gamma - \frac{7}{3}\right )^2 - \frac{3 \pi^2}{32} + \frac{19}{24} \notag  \\
& + \int_{-\infty}^\infty \frac{d u d v}{4\pi \sqrt{3}} \,e^{-2(u^2+u v+ v^2)/3}\, u^2 \ln(u^2)\, v^2 \ln(v^2)    \notag \\
& \approx  -0.039 .
\end{align}
Equation~\eqref{a2S} demonstrates that the result~\eqref{chi_fin_f} (obtained by expansion in powers of the correlation function~\eqref{DefCorr_0}) is valid provided the condition $y/(\bm{\beta}\pi^2)\ll 1$ holds.

\section{Asymptotic expression of the function $\mathcal{F}(x,y)$ at $y\gg 1$ \label{Appendix_MathF}}

In this appendix we outline derivation of the asymptotic expression~\eqref{tF_assympt} of the function $\mathcal{F}(x,y)$ at $y\gg 1$.  Provided $(2n+1)\leqslant |x|<(2n+3)$ with integer $n\geqslant 0$, the function $\mathcal{F}(x,y)$ can be rewritten as follows:
\begin{align}
\mathcal{F}(x,y) &=  \frac{1}{2} e^{-y/4} e^{y x/2} e^{-y x^2/4} \Biggl \{ \widetilde{\mathcal{F}}(x,y )
\notag \\
&+ 2 \sum_{m=0}^n  (-1)^{m}\exp\left [ y \frac{(|x|- (2m+1))^2}{4}\right ]\Biggr \} . \label{mathF_app}
\end{align}
Here we introduce the function
\begin{equation}
\widetilde{\mathcal{F}}(x,y) =\int_{-\infty}^\infty dt\, e^{-y t^2}\frac{\cosh(\pi t) \cos (\pi x/2)}{\sinh^2(\pi t)+\cos^2(\pi x/2)} ,
\end{equation}
which obeys 
\begin{equation}
\widetilde{\mathcal{F}}(x+2 k,y) = (-1)^{|k|} \widetilde{\mathcal{F}}(x,y),\quad \widetilde{\mathcal{F}}(2k+1, y)=(-1)^|k| ,
\end{equation}
for integer $k$. At $y\gg 1$ we find 
\begin{align}
\widetilde{\mathcal{F}}(x,y)& = \frac{\sgn [ \cos(\pi x/2)]}{y^{1/4}\sqrt{\cos(\pi x/2)}} \exp \left (\frac{y \cos^2(\pi x/2)}{2 \pi^2} \right ) \notag \\
& \times W_{-\frac{1}{4},\frac{1}{4}} \left (\frac{y \cos^2(\pi x/2)}{\pi^2}\right ) ,\label{Gassymp}
\end{align}
where $W_{\lambda,\mu}(z)$ denotes the Whittaker function. However, the Whittaker $W_{-1/4,1/4}(z)$  function is related with the error function:
\begin{equation}
W_{-\frac{1}{4},\frac{1}{4}}(z) = \sqrt{\pi} z^{1/4} e^{z/2} \Bigl [ 1-\erf(\sqrt{z})\Bigr ] .
\end{equation}
Therefore, for $y\gg 1$ we obtain 
\begin{align}
\widetilde{\mathcal{F}}(x,y) &= \sgn [ \cos(\pi x/2)]  \exp\left (\frac{y \cos^2(\pi x/2)}{\pi^2}\right ) \notag \\
& \times 
\left [ 1 - \erf\left (\frac{\sqrt{y}|\cos(\pi x/2)|}{\pi}\right )\right ] . \label{mathTF}
\end{align}
Using Eqs~\eqref{mathF_app} and \eqref{mathTF}, we find Eq.~\eqref{tF_assympt}.


\end{document}